\documentclass[useAMS,usenatbib]{mnras}

\usepackage{graphicx}
\usepackage{color}
\usepackage{multirow}
\usepackage{amssymb,amsmath}
\usepackage{cprotect}
\usepackage{makecell}

\usepackage[english]{babel}
\usepackage{blindtext}
\def\aap{A\&A}
\def\apj{ApJ}
\def\araa{ARAA}
\def\apjl{ApJL}

\def\pasp{PASP}
\def\pasa{PASA}
\def\mnras{MNRAS}

\def\na{New Astron.}

\def\apjs{ApJS}

\def\ps{$\rm km \,s^{-1}\,kpc^{-1}$}

\def\arcdeg{$^\circ$}
\def\kms{\,km\,s$^{-1}$}

\def\H2{$\rm H_2$}

\def\cc{$\rm cm^{-3}$}

\title[ISM and GMCs in different arm models]{How do different spiral arm models impact the ISM and GMC population?}

\author[A. R. Pettitt et al.]
{Alex R. Pettitt$^{1}$\thanks{E-mail:
alex@astro1.sci.hokudai.ac.jp},
Clare L. Dobbs$^2$,
Junichi Baba$^3$, 
Dario Colombo$^4$,\and
Ana Duarte-Cabral$^5$, Fumi Egusa$^6$ and Asao Habe$^1$
\\
$^{1}$Department of Physics, Faculty of Science, Hokkaido University, Sapporo 060-0810, Japan\\
$^{2}$School of Physics and Astronomy, University of Exeter, Stocker Road, Exeter, EX4 4QL, UK\\
$^3$National Astronomical Observatory of Japan, Mitaka, Tokyo 181-8588, Japan\\
$^4$Max-Planck-Institut f\"ur Radioastronomie, Auf dem H\"ugel 69, 53121 Bonn, Germany\\
$^5$School of Physics \& Astronomy, Cardiff University, Queen's building, The parade, Cardiff, CF24 3AA, U.K.\\
$^6$Institute of Astronomy, University of Tokyo, Mitaka, Tokyo 181-0015, Japan\\
}

\begin{document}
\date{\today}

\pagerange{\pageref{firstpage}--\pageref{lastpage}} \pubyear{2020}
\maketitle

\label{firstpage}

\begin{abstract}
{
The nature of galactic spiral arms in disc galaxies remains elusive. Regardless of the spiral model, arms are expected to play a role in sculpting the star-forming interstellar medium. As such, different arm models may result in differences in the structure of the interstellar medium and molecular cloud properties. In this study we present simulations of galactic discs subject to spiral arm perturbations of different natures. We find very little difference in how the cloud population or gas kinematics vary between the different grand-design spirals, indicting that the interstellar medium on cloud scales cares little about where spiral arms come from. We do, however, see a difference in the interarm/arm mass spectra, {and minor differences in tails of the distributions of cloud properties} (as well as radial variations in the stellar/gaseous velocity dispersions). These features can be attributed to differences in the radial dependence of the pattern speeds between the different spiral models, and could act as a metric of the nature of spiral structure in observational studies.
}
\end{abstract}


\section{Introduction}

The star formation process in a galaxy is inherently coupled to the structure of the interstellar medium (ISM) and the resulting population of giant molecular clouds (GMCs). The conditions of formation, evolution and eventual dissipation of these clouds is likely influenced by both cloud and galactic scale processes.

Many observational efforts hint at the properties of GMCs changing from galaxy to galaxy and within a given galaxy itself (e.g. \citealt{2018ApJ...860..172S,2013ApJ...779...46H,2016ApJ...822...52R,2020A&A...633A..17B}; Duarte-Cabral et al. subm.). The grand design spiral M51 in particular has garnered much attention from observational efforts \citep{2009ApJ...700L.132K,2012ApJ...761...41K,2017ApJ...836...62S}. \citet{2014ApJ...784....4C} discern a change in GMC properties between arm and inter-arm regions, with arms associated with clouds of higher masses and velocity dispersions. Recently \citet{2018PASJ...70...73H} performed a similar analysis to the barred-spiral M83, also finding clouds properties that change in different regions of the disc (such as the bar, arm, inter-arm regions).

Numerical works are also a powerful tool in understanding the properties of GMCs \citep{2015ApJ...801...33T,2019MNRAS.485.4997D}, with contemporary efforts pushing towards an understanding of cloud evolution across cosmic time \citep{2019arXiv191105251B,2020MNRAS.492..488G} and down to sub-pc scales \citep{2020MNRAS.492.1594S}. Exactly how clouds depend on the non-axisymmetric spiral structure has been the subject of only a few studies. \citet{2018MNRAS.475...27N} found that the impact of spiral features vastly exceeds changes possible from differing rotation curves (i.e. differing shear rates). \citet{2017MNRAS.464..246B} investigated the role of different types of spiral arms in terms of the GMCs population (simulating both a rigidly rotating and differentially rotating spiral pattern). They found that despite using two very different types of spiral arms, the global GMC population is effectively the same. \citet{2018MNRAS.480.3356P} performed simulations of an originally flocculent galaxy with arms then tidally triggered, and found that the GMC population transforms into one that is looser bound and with higher masses. In general, non-axisymmetric features seem to play a significant role in building up dense gas reservoirs that promote the formation of massive GMCs \citep{2011MNRAS.417.1318D,2014MNRAS.439..936F}.

The problem of nailing-down the fundamentals of spiral arm structure in galaxies has plagued the field for many decades (see \citealt{2014PASA...31...35D} for a review). The canonically accepted spiral model is that of a long-lived spiral density wave that rotates at some well defined pattern speed \citep{1964ApJ...140..646L}, which has been the subject of a great deal of follow-up works since its inauguration (see the review of \citealt{2016ARA&A..54..667S}). However, this model has been brought into question in recent years, mostly due to the results of numerical $N$-body simulations \citep{1984ApJ...282...61S,2012MNRAS.426..167G,2013ApJ...763...46B,2015MNRAS.449.3911P}. While the exact nature of such spirals is still debated, they manifest as a combined pattern that is dynamic, transient and recurrent in nature, a possible result of the superposition of a number of longer-lived yet still transient individual spiral modes \citep{2014ApJ...785..137S,2019MNRAS.489..116S}. An additional  possible progenitor for the observed spiral structure of galaxies is the interaction of some passing companion. Tidal forces exerted in such interactions can readily create 2-armed spiral features \citep{1972ApJ...178..623T,2011MNRAS.414.2498S,2018MNRAS.474.5645P} and such a mechanism is likely to have played a role in some of the more well-known spiral galaxies such as M51 \citep{2010MNRAS.403..625D} and M81 \citep{1999IAUS..186...81Y}. Bars can also drive spiral features in discs, particularly in the gas \citep{2001PASJ...53.1163W,2020MNRAS.491.2162P}, though for this study we limit ourselves to bar-free systems where the origin of spirality is more ambiguous. 

These different spiral theories all solicit different responses in the gaseous ISM. It is generally assumed that classical, quasi-stationary, spiral density waves rotate as solid bodies with some near-constant pattern speed, which results in gas moving in and out of the spiral perturbation at a rate that varies with galactic radius (due to flat galactic rotation curves). The passage of gas through these spiral arms is believed to induce shocks and other small scale structures such as interarm spurs and branches \citep{1969ApJ...158..123R,1968IAUS...29..453F,1973ApJ...183..819S}. These arms have a special co-rotation radius where gas and stars effectively rotate at the same speed as the spiral wave. In comparison, the spirals formed in live stellar discs instead appear to co-rotate with the gas at all radii, with gas effectively slowly falling into an arm from both convex and concave sides in large-scale colliding flows with no strong shock features \citep{2008MNRAS.385.1893D,2011ApJ...735....1W}. Spiral arms created in tidal interactions rotate as kinematic density waves (\citealt{1987gady.book.....B}, Chapter 6) that wind up at a rate somewhat slower than the material winding speed. Gas travels though the spiral perturbation but moves through solely from the concave side of the arms and at a similar rate throughout the disc \citep{2008ApJ...683...94O,2011MNRAS.414.2498S}, which can result in similar shocks and interarm features as seen in steadily rotating density waves \citep{2017MNRAS.468.4189P}.

Several previous works have attempted to draw out the nature of spirals from galaxies, usually by comparing numerical models to observational data. The Milky Way galaxy is an excellent laboratory for studying the nature of spiral (and bar) features, especially in the post-\emph{Gaia} era. A number of studies have tried to gauge the impact of spiral theories on stellar velocity fields in the Milky Way \citep{2014MNRAS.443.2757K,2015MNRAS.453.1867G,2018ApJ...853L..23B,2019MNRAS.484.3154S}, with a more transient spiral model seeming to reproduce features better. In addition, the ISM is believed to be very sensitive to changes in the galactic potential and it offers some of the more promising avenues to disentangling the nature of spiral arms. The work of \citet{2014arXiv1406.4150P,2015MNRAS.449.3911P} also found that the Milky Way's emission features was better fit by a more dynamic spiral pattern (see also \citealt{2018MNRAS.481.3794H}).

Outside of the Milky Way, a number of studies have also targeted extragalactic spirals, both through simulations and observations. A number of metrics of spiral arms have been suggested in the past, such as the age dating of clusters \citep{2010MNRAS.409..396D}, and the time-evolution of spiral pitch angles \citep{2019MNRAS.487.1808M}. Recent simulations have postulated a few key differences between the different spiral models mentioned in previous paragraphs. Such promising  metrics of spiral structure include: the radial dependence of spiral pattern speeds \citep{2008ApJ...688..224M}, gas streaming motions \citep{2016MNRAS.460.2472B}, spiral arm spurs \citep{2016MNRAS.458.3990P}, and radial profiles of offsets between gaseous and stellar material \citep{2015PASJ...67L...4B}. While some more broader comparative work exists that tests these facets across multiple models \citep{2019ApJ...876....6M,2019MNRAS.484.3154S}, the overall picture is far from complete, with evidence both for \citep{2018ApJ...869...29Y,2019ApJ...887...49S} and against \citep{2011ApJ...735..101F,2015ApJ...810....9C,2019MNRAS.490.1470P} the quasi-stationary density wave picture.

The aims of this paper are two-fold. Firstly we analyse how the GMC population varies between different models of spiral arms using numerical simulations, focusing on grand design 2-armed spiral galaxies. This is a fairly straight-forward question and will aid in inferring the nature of spiral features in future high resolution radio surveys. A secondary aim is to address how different spiral models leave their mark on the stars and gas in a more general sense, whether or not any change is seen in the GMCs. This study is similar in vein to \citet{2017MNRAS.464..246B}, who studied the impact of both rigidly rotating spiral features and those inherent to $N$-body simulations on GMC properties. We expand on this by assessing additional spiral generation mechanisms, and going a step further by looking in detail at differences in cloud properties in different models.

This paper is organised as follows. Section 2 describes the setup of the numerical simulations and the cloud identification technique. In Section 3 we present our results with a discussion of our findings in Section 4 and we conclude in Section 5.

\section{Methodology}
For this study we assess the role of different kinds of spiral arms in sculpting the star-forming ISM via numerically simulated galaxies where we identify if any tell-tale signatures could lie hidden in observational cloud populations. 
We use the same underlying galaxy model with the aim of a consistent cloud and ISM comparison, changing only what is necessary to vary the mechanism for forming spiral arms. We consider three different spiral models: a tidally induced 2-armed structure, a spiral generated by the self-gravity of the disc alone, and a spiral seeded by a rigidly rotating perturbation representing something akin to a density wave-like spiral.

\subsection{Simulation setup}

The models are based on the smoothed particle hydrodynamical disc galaxy simulation from \citet{2017MNRAS.468.4189P}, hereafter P17, performed using the \textsc{gasoline2} code\footnote{Publicly available at: \url{https://gasoline-code.com}.} \citep{2017MNRAS.471.2357W}. Four different galaxy models are used, three containing a grand design spiral structure. For the fourth we use a flocculent/multi-armed benchmark: the isolated disc of P17 which forms a many-armed bar-free disc (isolated with a small disc: IsoS). For the tidally generated spiral we use the main model from P17, generated by an in-plane prograde fly-by of a companion with IsoS (perturbed galaxy: Pert). The companion is gas-free and has a mass of 10\% of the primary galaxy, and reaches a closest approach on it parabolic orbit of 20\,kpc from the primary's galactic centre before moving away. To represent a spiral generated by disc instabilities we modify the IsoS setup to encourage the growth of low arm numbers of greater strength. To which end, the mass model was modified by increasing the disc to halo mass ratio to 50\% to encourage the growth of lower $m$ features compared to IsoS, where the galaxy has $m$-fold rotational symmetry, while keeping the same rotation curve shape (isolated with medium disc: IsoM). The final model is the IsoS disc subjected to the two-armed steadily rotating potential of \citet{2002ApJS..142..261C} used to represent a classical density wave style spiral (DenW). The potential is grown adiabatically over 200\,Myr and has a pitch angle of $15$\arcdeg and a pattern speed of 20\,\ps{}, similar to Milky Way values \citep{2011MSAIS..18..185G,2012MNRAS.425.2335S,2015MNRAS.449.2336J}. The resulting co-roation radius is 11\,kpc (very near the disc ``edge") and the inner-Lindblad resonance is around 4\,kpc.

Each model consists of an identical gas disc and bulge, with the stellar disc and halo being the same in three out of four (being slightly altered in the IsoM model). The initial gas particle mass resolution is 2000$M_\odot$ for all models, and the total gas mass is $6\times10^{9}M_{\odot}$. The gas surface densities and rotation curves are initially near-identical, and change only slightly with evolution. The rotation curves are all effectively flat outside of 2\,kpc and the gaseous surface density follows an exponential profile (see Appendix\;\ref{AppA} Figure\;\ref{VcSd}). The initial conditions themselves were generated using the \textsc{magalie} generator in the \textsc{nemo} toolbox \citep{1995ASPC...77..398T,2001NewA....6...27B}. The simulations were performed with sub-grid numerical recipes identical to P17, including a stochastic star formation from cold, dense gas ($T<300$\,K and $n>100$\,\cc{}) with a 6\% efficiency, tabulated cooling and heating processes \citep{2010MNRAS.407.1581S} with a fixed FUV background of $2.6\times 10^{-26}{\rm erg\,s^{-1}\,cm^{-3}}$ and blastwave-style supernova feedback \citep{2006MNRAS.373.1074S}. Self-gravity is active for all galactic components using a universal softening length of 10\,pc that is constant throughout the simulation.

\begin{figure}
\begin{centering}
\includegraphics[trim =10mm 0mm 10mm 0mm,width=85mm]{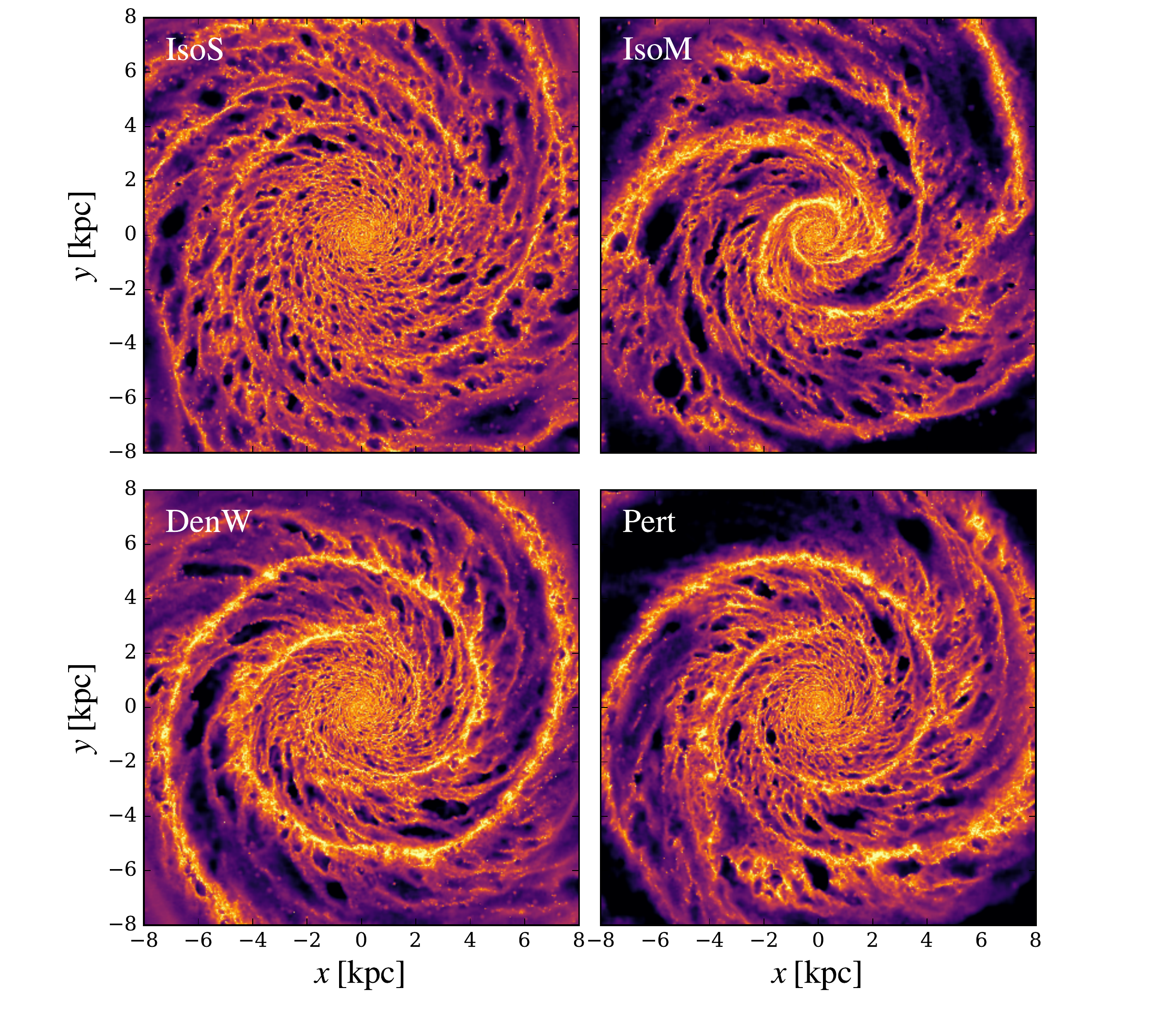}
\caption{Comparison of the morphology of the same galaxy with spiral arms generated by different mechanisms. 
These include dynamic spiral arms inherent to the original disc in isolation (top left), when the disc to halo mass ratio is increased (top right), those induced by a rigidly rotating density wave (bottom left) and a tidal passage of a satellite galaxy (bottom right).
}
\label{denmap}
\end{centering}
\end{figure}

We use a definition of spiral arm strength similar to that used in \citet{2015PASJ...67L...4B} to check the strength of the spiral perturbation:
\begin{equation}
\mathcal F(R) = \left.\frac{|\Phi -\langle\Phi\rangle_\phi|_{\rm max}}
{\langle\Phi\rangle_\phi}\right.
\frac{m}{ \sin \alpha}
\label{SpiralF}
\end{equation}
which quantifies the strength of the spiral as the variation in the radial forcing at a given radius. $\Phi$ is the gravitational potential at a given position, and $\langle \Phi \rangle_\phi$ is the mean over a specific annulus. This is averaged over 2 to 8\,kpc, and is given in Table\;\ref{modelvalues}. We also include the additional spiral potential term for the DenW model. The parameters $\alpha$ and $m_{\rm max}$ are the pitch angle and arm number of logarithmic spiral features fitted to the gas in the 2--8\,kpc range in each model using Fourier techniques, as described in \citet{2015MNRAS.449.3911P}. Specifically $m_{\rm max}$ is the mode number of the dominant Fourier amplitude, $A_m$, in the given radial range ($m$ varies as a function of radius for spirals driven by gravitational instabilities). The strengths of each grand design model are quite close, whereas the IsoS model is around 10\% weaker than the weakest grand design model. While $m_{\rm max}=3$ for IsoS, this mostly comes from the outer edge of the disc (see Fig.\;\ref{denmap}), while the inner disc has a much weaker signal than the models.

\begin{table}
\centering
 \begin{tabular}{@{}l | c c c c }
  Model & Note & $ m_{\rm max}$ & $\alpha\,[^\circ]$ & $\bar {\mathcal F}$ \\
  \hline
  IsoS & Isolated, multiple weak arms & 3 & 10.8 & 0.085\\
  IsoM & Isolated, few strong arms & 2 & 23.2 & 0.100 \\
  DenW & IsoS + rigid spiral potential & 2 & 12.4 & 0.098\\
  Pert & IsoS + perturbing companion & 2 & 11.2 & 0.094\\
 \end{tabular}
 \caption{Summary of the four galactic models. $m_{\rm max}$ is the arm number corresponding the the dominant Fourier mode, $\alpha$ is the pitch angle determined by a logarithmic spiral fit, and $\bar {\mathcal F}$ is the strength of the spiral (Eq.\,\ref{SpiralF}).  Note the spiral fitted to the gaseous material for DenW differs slightly to the additional underlying analytic potential, and the $\alpha$ above is the average of the two.}
 \label{modelvalues}
\end{table}

All models have been allowed to evolve for about 2 galactic rotations (500Myr) before analysis. For the IsoM model, we selected a specific time-frame that closely resembled an $m=2$ spiral. $N$-body spiral arms tend to be somewhat dynamic in nature, and can exhibit many different spiral modes within a single galactic rotation (e.g. \citealt{2012MNRAS.426..167G,2013ApJ...763...46B,2014ApJ...785..137S}) this disc in particular seemed to switch between a predominantly 2 and 3 armed morphology. Also note that such discs also favour lower arm numbers in the inner disc than the outer disc (which is usually attributed to a swing amplification effect), which is the opposite to what is seen in tidal fly-by spirals. 
 
\begin{figure}
\begin{centering}
\includegraphics[trim =0mm 10mm 0mm 0mm,width=85mm]{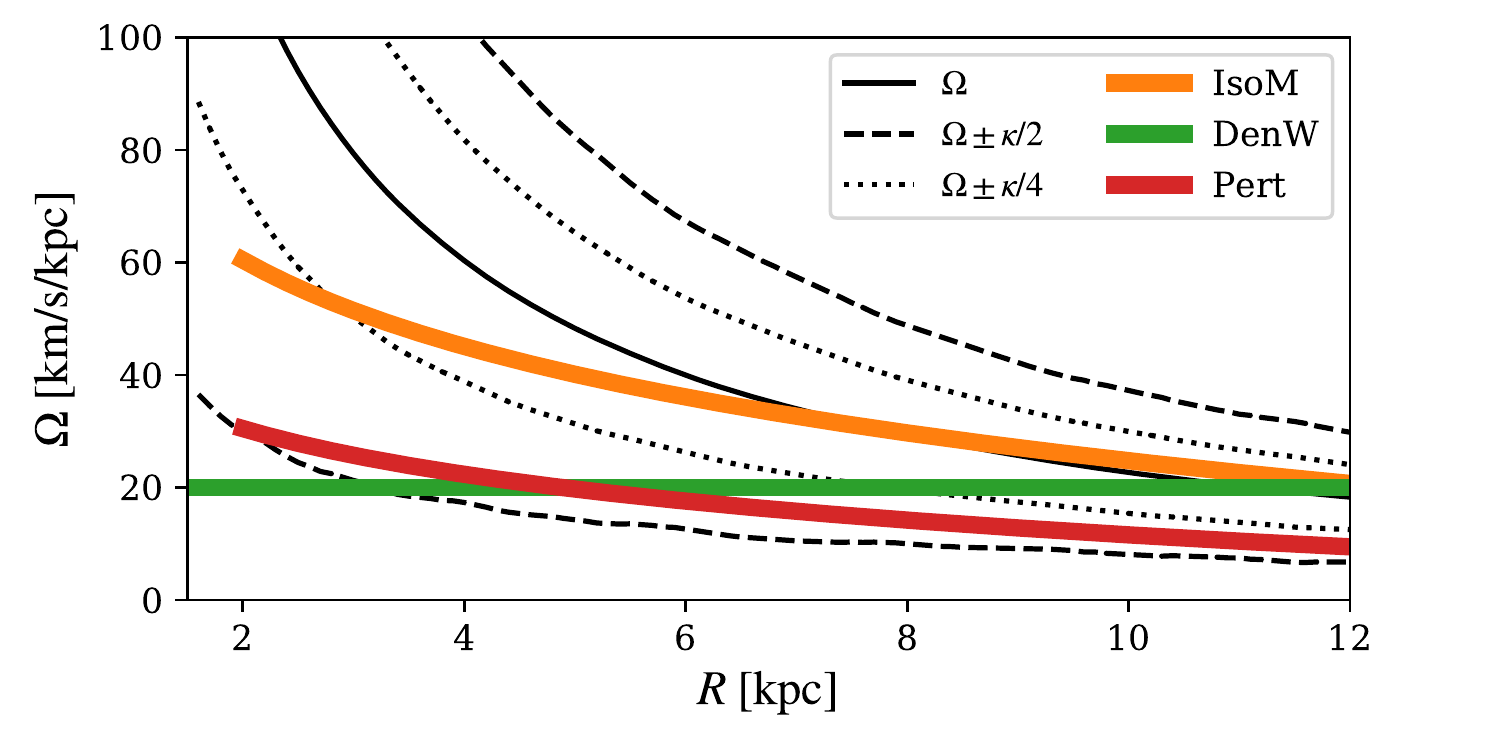}
\caption{Arm pattern speeds as a function of radius in each of the grand design spiral models shown alongside the frequencies of the disc material. $\Omega$ and $\kappa$ are the material and epicycle frequencies of the gas disc, respectively.}
\label{pspeeds}
\end{centering}
\end{figure}

Note that behaviour of the pattern speed is different for all models, and is shown in Figure\;\ref{pspeeds} for each grand design spiral model. To calculate the pattern speed for IsoM and Pert spiral arms are fit to the gas structure 10\,Myr prior to the timeframe shown in Figure\;\ref{denmap}, and the azimuthal displacement between those two measurements gives pattern speed as a function of radius. The Pert model has a pattern speed of $30-5$\,\ps{}, decaying with radius at a rate similar to $\Omega-\kappa/2$ where $\Omega$ is the gas rotation frequency and $\kappa$ the epicycle frequency (see also \citealt{2016MNRAS.458.3990P}). The spiral pattern exhibited by IsoM (and also IsoS) appears to be closer to the material speed of the disc, though is inherently a superposition of many concurrent modes \citep{2014ApJ...785..137S}. Towards the disc centre ($R<4$kpc) this flattens out, possibly betraying the existence of a growing bar-mode that is yet to manifest, though it is also common in other simulation works for pattern speeds to flatten out at values $<\Omega$ in the inner disc (e.g. figure 3 of \citealt{2015PASJ...67L...4B}, figure 4 of \citealt{2012ApJ...751...44S}). The DenW arm potential rotates at 20\,\ps{}, which is a parameter set in defining the potential. While this could be decreased to fit more in line with the Pert model in the mid/outer disc, it would result in the arm response moving out to greater radii in accordance with the expansion of the inner Lindblad resonance, which would make comparisons between arm models even more difficult.

\subsection{GMC definition}
We use the exact same extraction method and parameters as in \citet{2018MNRAS.480.3356P}, which details the extraction of the cloud catalogue from the Pert and IsoS discs. In brief, clouds are extracted using a `friends-of-friends' approach \citep{2015MNRAS.446.3608D}. A cut is density is first made, only considering GMC particles above $40$\cc{}, and a neighbour search around each dense gas particle is performed over a distance of 15\,pc. A mass resolution limit of $8\times 10^{4}{\rm M_\odot}$ is imposed for the clouds, requiring at it contains at least 40 particles. The clouds masses, $M_{\rm c}$ are simply the sum of particle masses defining a cloud. The radius, $R_{\rm c}$, is defined by mean surface area projected in three orthogonal planes \citep{2014MNRAS.439..936F}. The velocity dispersion of the clouds, $\sigma_{\rm c}$, is defined using the variance of the 3D motion about the centre of mass of the cloud. The virial parameter is calculated from these properties as: $\alpha_{\rm vir}=5\sigma_{\rm c}^2 R_{\rm c}/GM_{\rm c}$, and we use the specific angular momentum in the $\hat z$ direction, $l_z$, to quantify the spin of the clouds.

We remind the reader that the choice of cloud definition and identification is very important when attempting to compare simulated clouds to catalogues from observations, or when comparing numerical or observational studies to each other (e.g. Fig. 15 of \citealt{2019MNRAS.483.4291C}). While this is something we will take into account in future works that include such comparisons, the single method adopted here is sufficient for determining the relative changes in the cloud population between a set of idealised numerical models.

The time-evolution of the cloud properties in IsoS and Pert were the topic of \citet{2018MNRAS.480.3356P} (e.g. their Fig. 6 and 7). It was seen that the cloud population stayed relatively consistent in both of the two different morphological states: the period immediately after the interaction (Pert), and before companion passage (IsoS). We will focus on our model galaxies at a single instance in this work, with the aim of seeing if any features can be teased out of a single snapshot of a galaxy that can shed light on the different types of spiral arms. This is a justified approach for the two new models. For DenW there is no time-dependence in the potential, and so once the spiral perturbation has reached full strength the gas remains the same. For IsoM the situation is somewhat trickier as the spiral arms are quite dynamic in nature, with the overall structure changing from a 2 to 3 armed spiral in the course of a dynamical time. This means any time-dependent analysis would have to contend with the morphology actively changing as well, removing the constraints of looking at systems will predominantly $m=2$ grand design structure.

\section{Galactic structure}

\subsection{Disc morphology}

Figure\;\ref{denmap} shows a top-down render of gas density for all four models considered. By-eye the three grand-design discs appear quite similar, and the IsoS model clearly has the weaker arm structure. While IsoM does appear to be primarily a two-armed spiral, there is a significant tertiary arm feature in $x>0$, $y>0$. Such bifurcation of arms is common to $N$-body spirals, with simulations showing a number of arms that increases with radius \citep{2015ApJ...808L...8D,2015MNRAS.449.3911P}. Long-lived two-armed arm features are known to be difficult to maintain in $N$-body discs \citep{2011MNRAS.410.1637S,2014PASA...31...35D}. Table\;\ref{modelvalues} lists the spiral arm number and pitch angle fitted via a Fourier analysis, indicating that the pitch angle of the IsoM model is significantly more open than the other models. This is also not uncommon in the literature, with low arm number spirals created in unbarred $N$-body discs with realistic rotation curves having similarly wide pitch angles \citep{2013A&A...553A..77G}. The spiral arms fitted to the DenW disc are tighter than the imposed analytic potential. This is as expected from theory, owing to gas shocking at an azimuth offset to the underlying spiral potential that varies with galactic radius \citep{1968IAUS...29..453F,1969ApJ...158..123R} and results in pitch angle for spiral arms that differs depending on the specific tracer. This has been observed in simulations \citep{2014MNRAS.440..208K,2015PASJ...67L...4B} and some observed galaxies \citep{2017MNRAS.465..460E,2018ApJ...869...29Y}.

The DenW and Pert discs show spurring features downstream of the arms, which do not appear in the unperturbed discs, as noted before in other works \citep{2006MNRAS.367..873D,2006ApJ...647..997S}. The exact nature of these spurs has been debated, being attributed to Kelvin-Helmholtz instabilities \citep{2004MNRAS.349..270W}, vorticity at the deformed spiral shock front \citep{2014ApJ...789...68K}, or crowding and subsequent shearing of accumulations of cold gas as it leaves the arms \citep{2006MNRAS.367..873D}. In \citet{2016MNRAS.458.3990P} the authors reported (their figure 9) that spurs differ between spiral models, utilising much lower resolution and assuming a warm, isothermal ISM. An in-depth look at the differences between spurs in different spiral models is being readied for a future study.

Additional properties of these discs are given in Appendix\;\ref{AppA}, showing very similar rotation curves and gas surface densities in each model.

\subsection{Disc kinematics}

\begin{figure*}
\begin{centering}
\includegraphics[trim =0mm 0mm 0mm 0mm,width=175 mm]{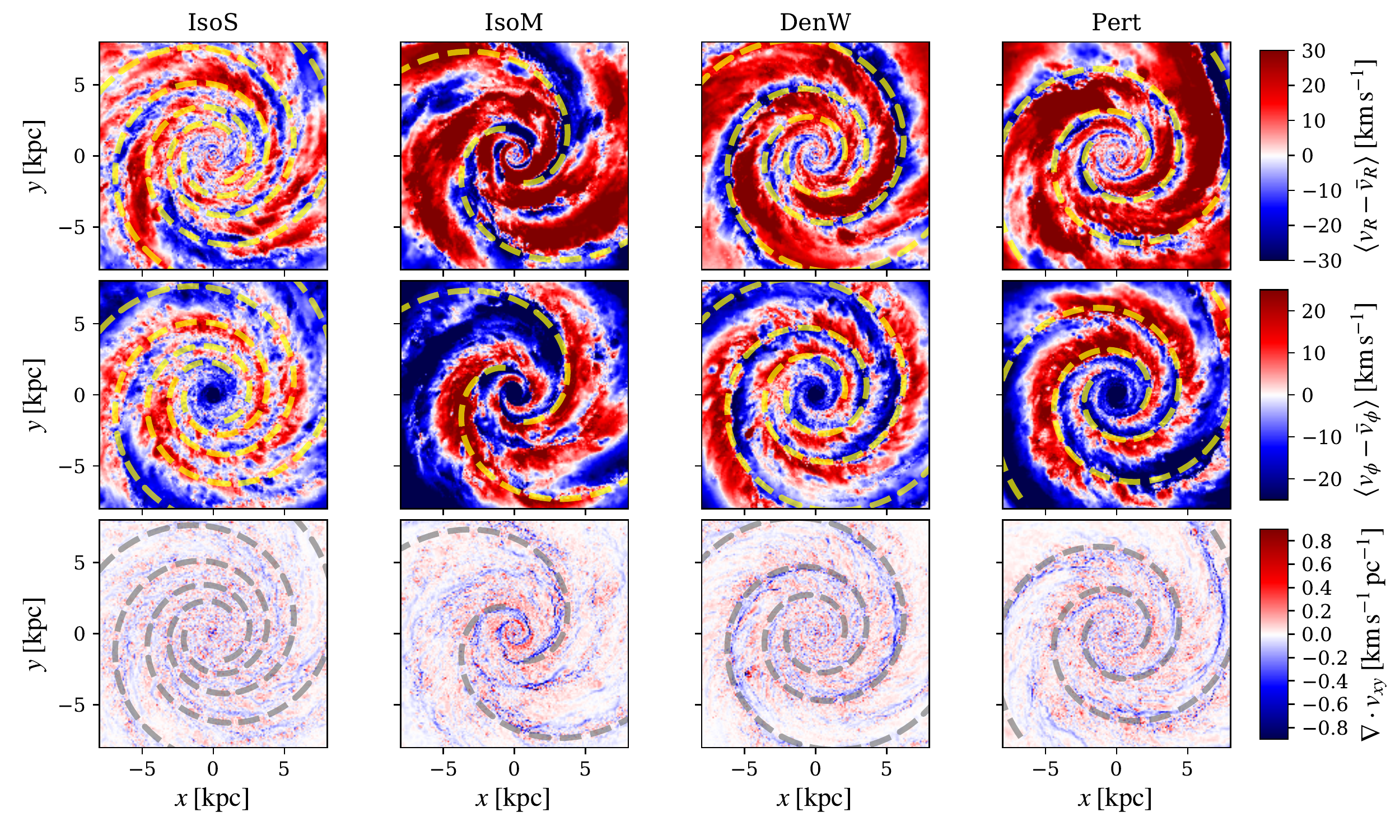}
\caption{The changes in non-circular motions for gas in each model galaxy, where quantities have been interpolated using the SPH smoothing kernel. Colour-scales indicate the  velocities in compared to the median value at a given radius. Spiral arm fits discussed in the text are over-plotted as dashed yellow lines. From top: radial streaming motion, azimuthal streaming motion, and the divergence of the 2-dimensional, in-plane, velocity field.}
\label{gasstream}
\end{centering}
\end{figure*}

In Figure\;\ref{gasstream} we show the streaming velocities in each of the models. The azimuthal and radial velocities are shown in the top and middle rows respectively. The colour indicates the difference from the median values in each given cell. While it would be expected for each disc to have a mean radial velocity component centred at 0\,\kms, the Pert disc has experienced a strong radial perturbing force (migrating from its initial centre-of-mass position), so there is an overabundance of outwards motion compared to the other models. This stronger radial signature could be taken to be a signature of tidally driven spiral arms, however, this should be done so with caution as the radial migration of gas is a product of the interaction strength and the evolutionary stage of the encounter (see figures 19 and 20 of \citealt{2016MNRAS.458.3990P}).

For each model the streaming motions are quite similar with respect to the spiral potential well, which is indicated by the overplotted dashed lines, and each grand design model displays similar magnitudes of non-circular motion. An interesting difference is that the spur features show up very clearly in DenW and Pert when viewing in $v_\phi$, and are areas where gas maintains a positive (red in the figures) azimuthal velocity. The lower row shows the divergence of the in-plane velocity fields ($\nabla \cdot v_{xy}$), which acts a converging flow criteria for the star formation recipe. In general there is no strong difference between the models, with each showing a correspondence between converging flows and the spiral arms. There is a slight preference for converging flows to start slightly upstream of the arm minima, but the density still remains too low at this point to trigger formation of stars until it enters the arm.

\begin{figure*}
\begin{centering}
\includegraphics[trim =0mm 0mm 0mm 5mm,width=160mm]{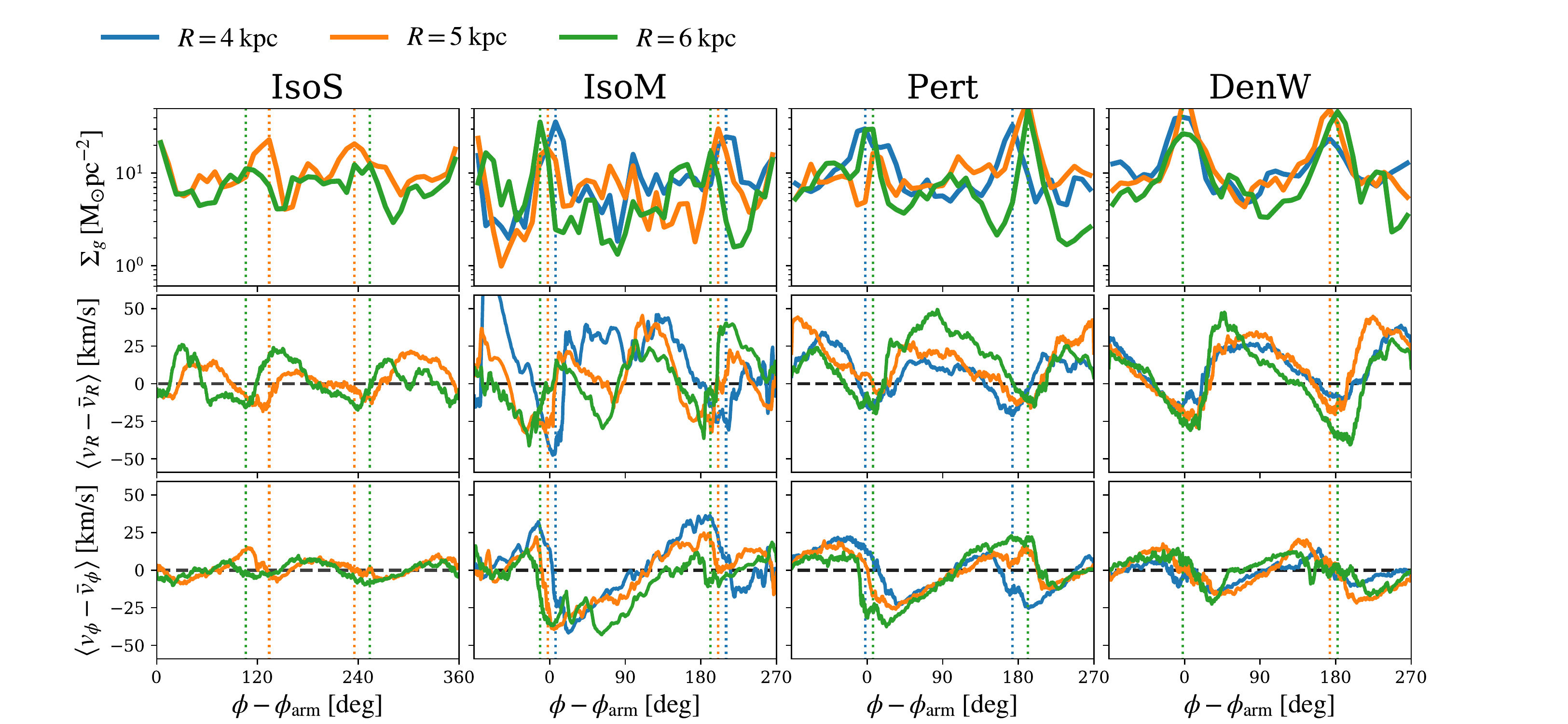}
\caption{ISM density and velocities in three concentric rings of 1\,kpc in width in four different armed models at three separate radii (colours). The $x$-axis indicates the azimuthal position of gas ($\phi$, measured anti-clockwise in Figure\;\ref{denmap}) shifted in azimuth to the azimuthal position of the fitted spiral model ($\phi_{\rm arm}$, in accordance with the log-spiral model in Table\;\ref{modelvalues}) so that each radius show the relative position of gas to the spiral arm potential. The three rows show large scale properties of the gas: the gas surface density, radial and azimuthal streaming motions. Gas moves from right to left in these panels. Vertical lines indicate the location of the peak gas density in each arm. Coloured vertical lines show the exact location of the peak gas density at each radius.}
\label{AzVels}
\end{centering}
\end{figure*}

In Figure\;\ref{AzVels} we again show the ISM streaming motions, but now as a function of azimuthal angle phase-shifted to the arm minima, $\phi-\phi_{\rm arm}$, where $\phi_{\rm arm}$ is the azimuth of the fitted log-spiral arm model. $\phi$ is defined positive anti-clockwise and so gas moves from right to left in these panels. The actual locations of the peaks in gas density are not precisely at this shifted location, due to the gas not perfectly matching a log-spiral fit at all radii. The top row shows the actual gas surface density, with density maxima marked as dashed vertical lines across all panels. No 4\,kpc data is shown for IsoS model, as there are no discernible spiral arms at this radius. The three grand design models clearly trace out a 2-armed pattern in surface density, with similar arm:inter-arm contrasts in each. The additional strong inter-arm feature in IsoM is located at around $-80$\arcdeg. IsoS instead favours 3 arms at these radii (the third is located at the panel edges). The second and third rows show the radial and azimuthal streaming motions. The velocity patterns in each of the grand design discs are remarkably similar at all radii, making it difficult to discern between the different models when comparing streaming motions alone. Even the IsoS model shows similar trends, though with much reduced amplitude, despite the arms being much weaker than in the other models. Spiral arms approximately line up with the troughs in $v_{R}$ and saddle points of $v_{\phi}$. This is qualitatively similar to what is seen in observations of the grand design spiral M51 in \citet{2007ApJ...665.1138S}.

Similar streaming motions in each grand design spiral is somewhat different to what was reported in \citet{2016MNRAS.460.2472B}, where the authors found that their dynamic arms (akin to IsoM) showed slightly different streaming motions compared to a static spiral potential (akin to DenW). We propose this difference may be attributed to two reasons. The first is that their model has an inner-bar component, which moves at around 40--50\,\ps{} (their Fig.\;3), which corresponds to an outer 2:1 (OLR) resonance in their model at 7--10\,kpc. It has been shown in numerous previous works that bars can influence the dynamics of gas out to the OLR (e.g. \citealt{2002A&A...396..867K,2020MNRAS.491.2162P}), which could change the trends in streaming motions in their analysis (performed between 6--8\,kpc) compared to what would be seen in a spiral-only model. Secondly, their arms are quite weak compared to the static spiral they use for comparison, with numerous inter-arm branches and wide density peaks. It may be that the gas streaming motions in the dynamic-style arms depend on the evolutionary stage of any given arm (i.e. are time-dependent), and so the data here and of \citet{2016MNRAS.460.2472B} may not be true for all such arms in general. A full analysis of streaming motions in dynamic-style arms as they grow and decay is beyond the scope of this work and we defer to a future study.

Dynamic-style spiral arms displaying similar streaming motions to more wave-like perturbations such as DenW and Pert is perhaps not too surprising if the arms form from a superposition of many underlying waves, each with its own pattern speed and resonance radii \citep{2014ApJ...785..137S}. In light of that interpretation, both the dynamic and static style arms in some way stem from some long-lasting wave features with fixed pattern speeds, and so the similarity in the gas response would be expected.

\section{Cloud analysis}

\subsection{Comparisons between the cloud populations}

\subsubsection{Properties of clouds}
\begin{table*}
\centering
\def\arraystretch{1.5}
 \begin{tabular}{@{}l | c c c c c c c c c c c c }
  Model & $N_{\rm c}$ & $\sum M_{\rm c}$ & $\sum M_{\rm c}/M_{\rm g}$ & $\gamma$  & $\langle M_{\rm c} \rangle$ & $\langle R_{\rm c} \rangle$ & $\langle \sigma_{\rm c} \rangle$   & $\langle \alpha_{\rm vir} \rangle$  & $\langle l_z^{\rm pro} \rangle$ & $\langle l_z^{\rm ret} \rangle$ & $p_{\rm IsoS}$   \\
      &   &  $[M_{\odot}]$ &  [\%] &    &  $[10^5 M_\odot]$  &   $[\rm pc]$ & $[\rm km\,s^{-1}]$ &  &  $\rm [pc\,km\,s^{-1}]$ &  $\rm [pc\,km\,s^{-1}]$  &   \\  
  \hline
  IsoS   & 923 & $2.1\times 10^8$ & 4.8  & $-2.58$ & $1.5^{+2.4}_{-1.1}$ & $30.8^{+37.5}_{-26.7}$ & $3.0^{+4.1}_{-2.2}$ &  $1.9^{+3.2}_{-1.1}$ &  $37.4^{+95.9}_{-14.6}$ &  $15.2^{+30.0}_{-5.5}$ & --\\
  IsoM   & 786 & $2.7\times 10^8$ & 7.4  & $-2.19$ & $1.7^{+3.2}_{-1.2}$ & $33.4^{+42.0}_{-28.6}$ & $3.8^{+5.6}_{-2.6}$ &  $2.6^{+5.7}_{-1.5}$ &  $40.6^{+122.2}_{-11.9}$&  $18.5^{+45.1}_{-7.2}$ & $10^{-4}$\\
  DenW & 1010 &  $2.9\times 10^8$ & 7.2  & $-2.37$ & $1.7^{+2.8}_{-1.1}$ & $32.4^{+40.0}_{-28.0}$ & $3.3^{+4.4}_{-2.4}$ &  $2.2^{+3.6}_{-1.3}$ &  $38.8^{+106.8}_{-14.7}$ &  $16.7^{+35.9}_{-7.6}$ & $10^{-2}$\\
  Pert   & 689 & $2.4\times 10^8$ & 7.7  & $-2.39$ & $1.8^{+3.6}_{-1.2}$ & $34.2^{+43.4}_{-28.6}$ & $4.0^{+5.9}_{-2.9}$ &  $3.0^{+5.4}_{-1.8}$ &  $58.9^{+178.2}_{-20.3}$ &  $15.4^{+43.5}_{-6.5}$ & $10^{-8}$\\
 \end{tabular}
 \caption{Summary of extracted clouds from each simulation, showing the total number of clouds, the mass of all clouds, the percentage of the gas budget of the galaxy allocated to clouds, and the fit to the slope of the mass spectrum. The uncertainty on $\gamma$ is $\pm 0.1$ across all models. Medians for each parameter shown in Figure\;\ref{ModelHistCompV} are given in the format $\langle X\rangle ^{+\rm Q75}_{-\rm Q25}$ where upper and lower indexes are lower and upper quartiles. The value $p_{\rm IsoS}$ is the $p$-value for the KS-test of masses compared to the IsoS model.}
 \label{tab1}
\end{table*}

The properties of the extracted cloud catalogues is given in Table\;\ref{tab1}. DenW has the largest number of clouds, though this is likely a product of the spiral wave extending across a large radial range compared to the other models (IsoM has strong central arms, Pert strong outer arms). Pert has the fewest number of clouds at this instance, though is not much different to IsoS, which is somewhat surprising given how flocculent that disc appears. The three models with low spiral arm numbers have a remarkably similar fraction of their gas allocated to clouds, all being around 7\%. Note this mass fraction is defined with respect to the total gas reservoir of the galaxy, whereas literature surveys usually present this as a fraction of molecular mass/flux, which is naturally more correlated with GMCs and so gives higher fractions (e.g. 55\% in M51 from \citealt{2014ApJ...784....3C} and 25\% in the Milky Way from \citealt{2016ApJ...822...52R}). The precise cloud definition technique used will also alter these fractions. Note that each disc has a different star formation history (see Fig.\;\ref{SFH}), so the total gas mass is different in each model at the epoch of GMC analysis.

\begin{figure}
\begin{centering}
\includegraphics[trim =0mm 10mm 0mm 0mm,width=85mm]{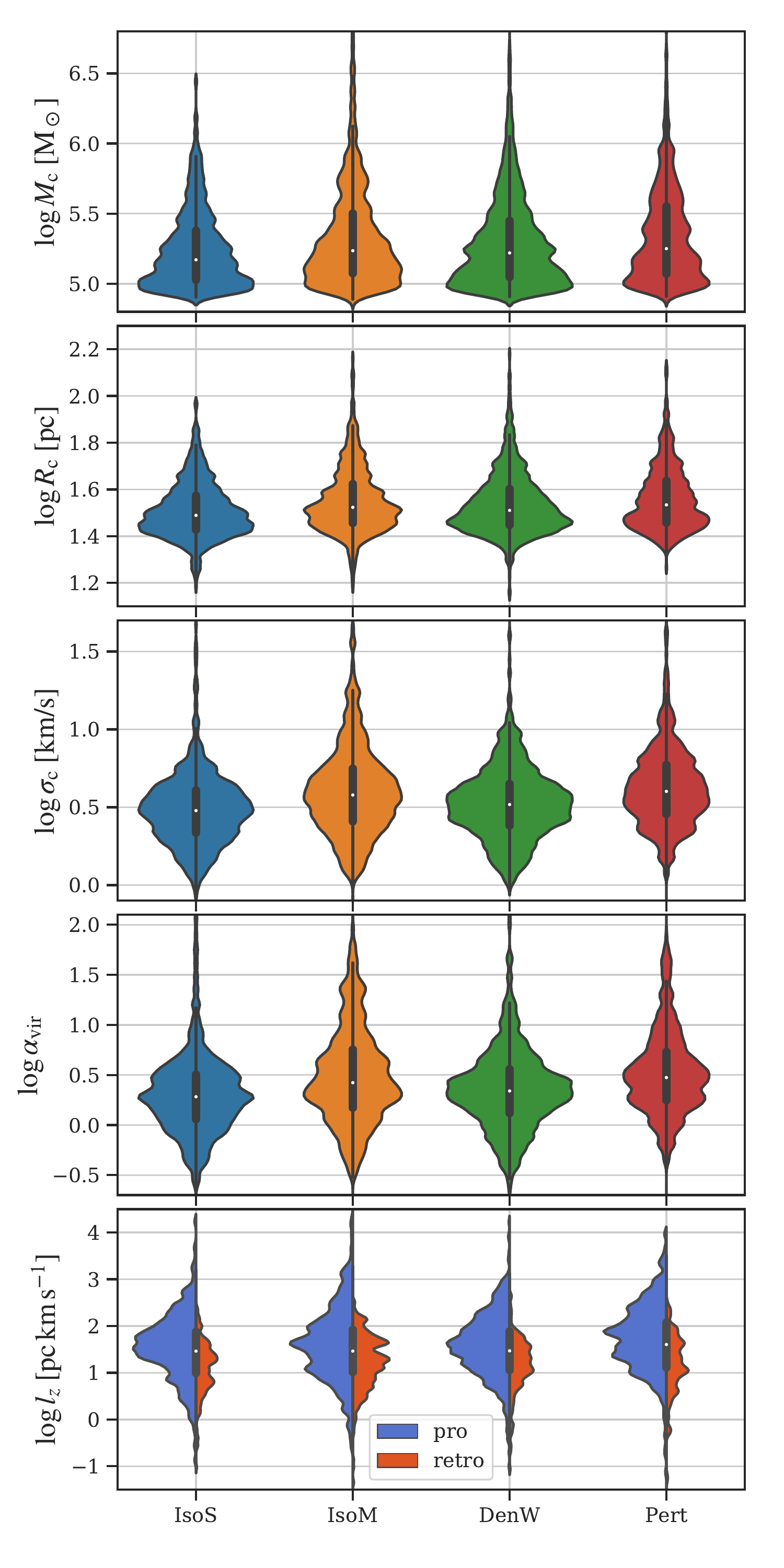}
\caption{Violin plot of various properties of GMCs in each different model using the entire disc population. Top to bottom we show the clouds mass, radius, velocity dispersion, virial parameter and angular momentum in the $z$-direction, the latter is shown for both pro- and retrograde spins. Inner box plots span the quartile ranges and whiskers extend to 1.5 the interquartile range. Histogram versions of the same data are shown in Appendix\;\ref{AppB}.}
\label{ModelHistCompV}
\end{centering}
\end{figure}

In Figure\;\ref{ModelHistCompV} we show violin plots of cloud masses, radii, velocity dispersions, virial parameters and specific angular momenta in each model ($M_c$, $R_c$, $\sigma_c$, $\alpha_{\rm vir}$ and $l_z$ respectively). The scales in the $x$-axes are the true number of counts in each bin, rather than normalised counts. The lowest panel splits the angular momenta into prograde (blue) and retrograde (red) clouds. Various statistical properties of these different cloud populations are given in Table\;\ref{tab1}, showing the same median and quartile ranges as indicated in the inner box plot of Figure\;\ref{ModelHistCompV}. 

While the overall shape of each distribution is similar, there are several subtle differences between the different GMC populations. The IsoS disc shows an inability to grow larger cloud structures, due to the spiral pattern being less grand design in nature compared to the clear two arms of the other models. All three grand design spirals have very similar high mass tails of their mass functions and size distributions. The main difference is that Pert (and to a smaller extend IsoM) seems to be relatively inefficient at maintaining a low mass and small radius cloud population. This could be a result of the continual merger of smaller cloud structures within arms in the tidal disc compared to the other models. It may also be in part due to the other models having cloud populations that extend to much larger galactic radii where many smaller clouds reside (see Fig.\;\ref{alphamap}) whereas the tidally perturbed model has had much of its outer disc stripped of gas by the companion. We will investigate the population more consistently in Sec.\;\ref{sec_arm2} by analysing a specific arm regions in each disc.

Similar trends are seen in the velocity dispersion and virial parameter of the clouds across all models. IsoS has the clouds with lowest values of $\sigma_c$ and $\alpha_{\rm vir}$, though this is not too dissimilar to DenW. IsoM seems to create slightly more unbound clouds, possibly a by-product of many of its clouds (and dense gas) being closer to the galactic centre compared to the other models, where the strong non-circular motions of bulge stars could act to disrupt the clouds more efficiently. The tidally perturbed disc has the largest dearth of clouds with low $\sigma_c$ and $\alpha_{\rm vir}$. The distribution of clouds in $l_z$ space shows only small deviations between models. The fraction of retrograde clouds shows no change between the models, which is somewhat surprising if the strong sweeping up of gas in the tidal model would be lowering the collision timescale between clouds. It may be the case that any non-axisymmetric perturbation will induce retrograde cloud rotation so long as feedback can break apart clouds before they experience runaway growth \citep{2011MNRAS.417.1318D}. The tidal interaction does produce a smaller fraction of slowly rotating prograde clouds (seen by the upwards shift of the box plot compared to other models), signifying the imparted angular momentum from the companion to the disc. We will look at cloud rotation in more detail in Section\;\ref{sec_rot}. 

\subsubsection{Statistical differences}

We performed a fit to the cumulative mass spectra of the clouds in each model and the resulting slopes\footnote{Given by $\gamma$ in the fit to the cumulative mass spectra of the form $N_0[(M_C/M_0)^{\gamma+1}-1]$ where $M_0$ and $N_0$ are effectively free parameters, see \citet{2005PASP..117.1403R} for details.} of the spectra ($\gamma$) are given in Table\;\ref{tab1}. An uncertainty on $\gamma$ is estimated from a resample-without-replacement statistical analysis of the cloud masses by sampling 30\% of the data over 100 trials, with the standard deviation on the resulting fitted $\gamma$ values providing an uncertainty of $\pm 0.1$ for all models. The values of $\gamma$ agree with what is visible by-eye, in that the more flocculent disc (IsoS) has a steeper slope due to the dearth of high mass clouds compared to the other models. The DenW and Pert models have nearly identical slopes, with flatter distributions extending to higher masses than IsoS. IsoM has the flattest distribution, though this is heavily influenced by a couple of clouds around $\log M_c=6.5$ which pull the distribution up significantly. The fitted values of $\gamma$ lie within error bars of each other for each of the grand-design spiral models, but outside those for IsoS. Using different sizes of bins has a slight impact on the fitted value of $\gamma$, but the trends between the models remain the same (IsoS is always the steepest and IsoM the shallowest).

A Kolmogorov-Smirnov (KS) test was performed on each population compared to IsoS and found that all models are statistically distinct compared to IsoS for all parameters. The distributions of $M_c$ tend to be the closest to the values of the IsoS population, with \mbox{$p$-values} approximately $10^{-4}$, $10^{-2}$ and $10^{-8}$ for IsoM, DenW and Pert respectively. While it is not really surprising that IsoS differs from the stronger arm models, the difference between IsoM, DenW and Pert is harder to distinguish from the figures. KS tests against those populations reveal that DenW and IsoM are very similar, with $p$-values across parameters ranging from 0.1 to 0.01, implying they are only marginally distinct from each other. DenW and IsoM have similar $M_c$ and $R_c$ distributions ($p$-values of 0.1 and 0.02 respectively) though the differences in $\sigma$ and $\alpha_{\rm vir}$ are much more significant ($p$-values around $10^{-8}$). In comparing Pert to DenW it is seen that both have fairly similar mass and radii distributions ($\sim10^{-3}$) but very different $\sigma_{\rm c}$ and $\alpha_{\rm vir}$ ($\sim 10^{-12}$). This effectively suggests the distribution of masses and radii are very similar in the strong spiral models, though they show differences in cloud velocity dispersions (which lead to differences in $\alpha_{\rm vir}$).

\subsection{Unbound clouds and disc dispersion}

\begin{figure}
\begin{centering}
\includegraphics[trim =10mm 0mm 0mm 0mm,width=95mm]{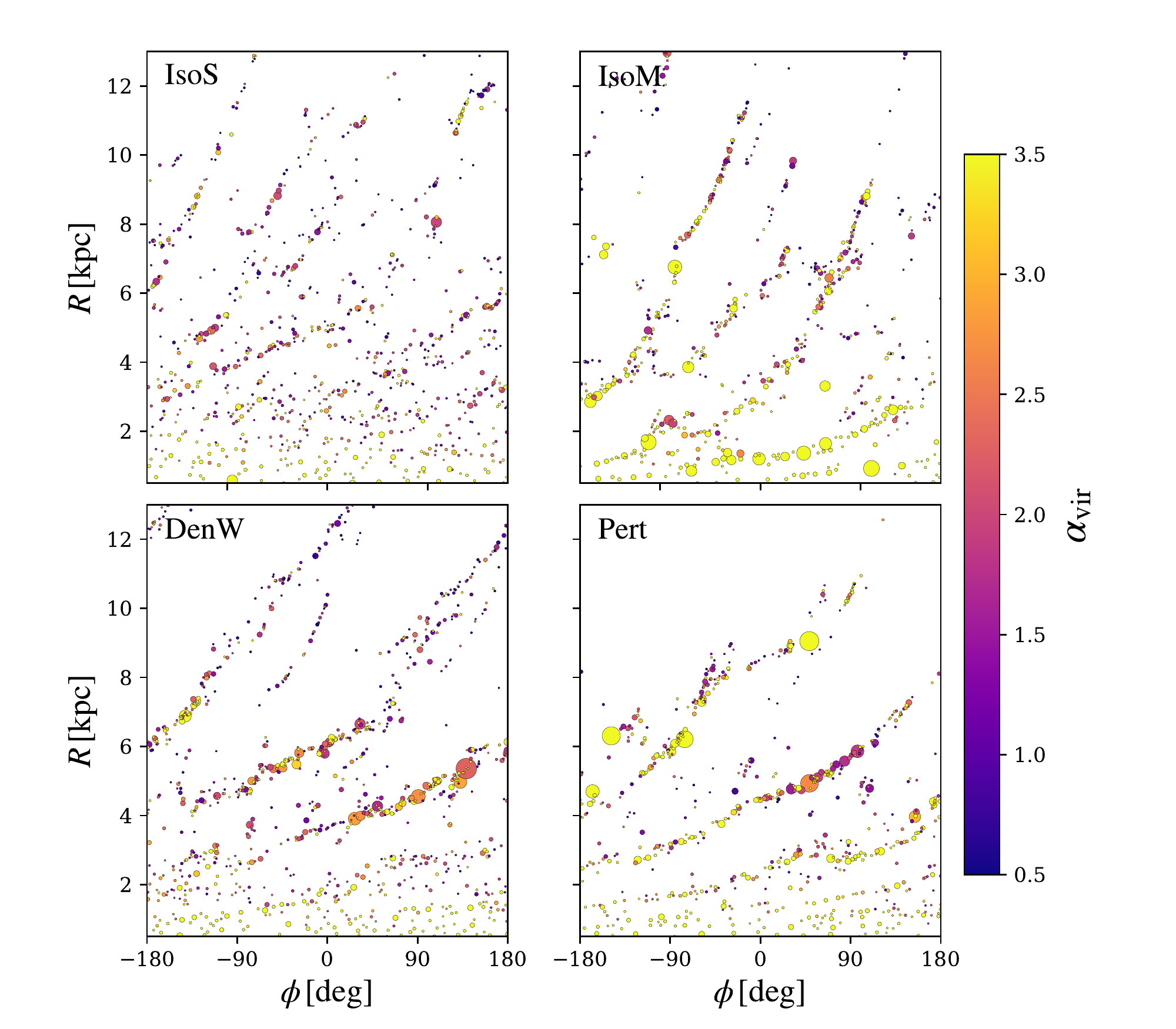}
\caption{Clouds extracted from the simulations projected in the radial-azimuthal plane. Clouds are coloured by virial parameter and sizes indicate relative mass.}
\label{alphamap}
\end{centering}
\end{figure}

In Figure\;\ref{alphamap} we show maps of locations of clouds in $R-\phi$ space. Clouds are coloured by $\alpha_{\rm vir}$ and sized by mass. The clouds in IsoS appear to be moderately more well-bound than the other models, though still trace out the weak spiral pattern. The three grand-design spirals have clouds clearly tracing the spiral arms. Pert and IsoM have a significant population of large and loosely bound clouds. DenW has a well-bound population at the edge of the disc, not present in the IsoS and Pert. The response to the spiral potential in the DenW model is quite limited at the disc edge. The co-rotation is near 11\,kpc for DenW, and while gas responds to a spiral potential between Lindblad radii, the response at co-rotation is often mild (e.g. \citealt{2014ApJ...789...68K}, where wiggle-instabilities from strong spiral shocks only occur away from co-rotation, where relative flow to the potential becomes sub-sonic). In this region of DenW the cloud population becomes more akin to that of IsoS, with some mild re-arrangement into an $m=2$ pattern caused by the gravitational torque of the occasionally passing potential. Pert and IsoM are quite similar in cloud distribution. IsoM has a higher degree of structure due to the additional features between the $m=2$ components. The least bound clouds are either relegated to inner regions of the disc (where disc shear is strongest) or are the most massive clouds. Well bound clouds seem to only populate small/moderate masses in the mid to outer disc.

\begin{figure}
\begin{centering}
\includegraphics[trim =0mm 0mm 0mm 0mm,width=70mm]{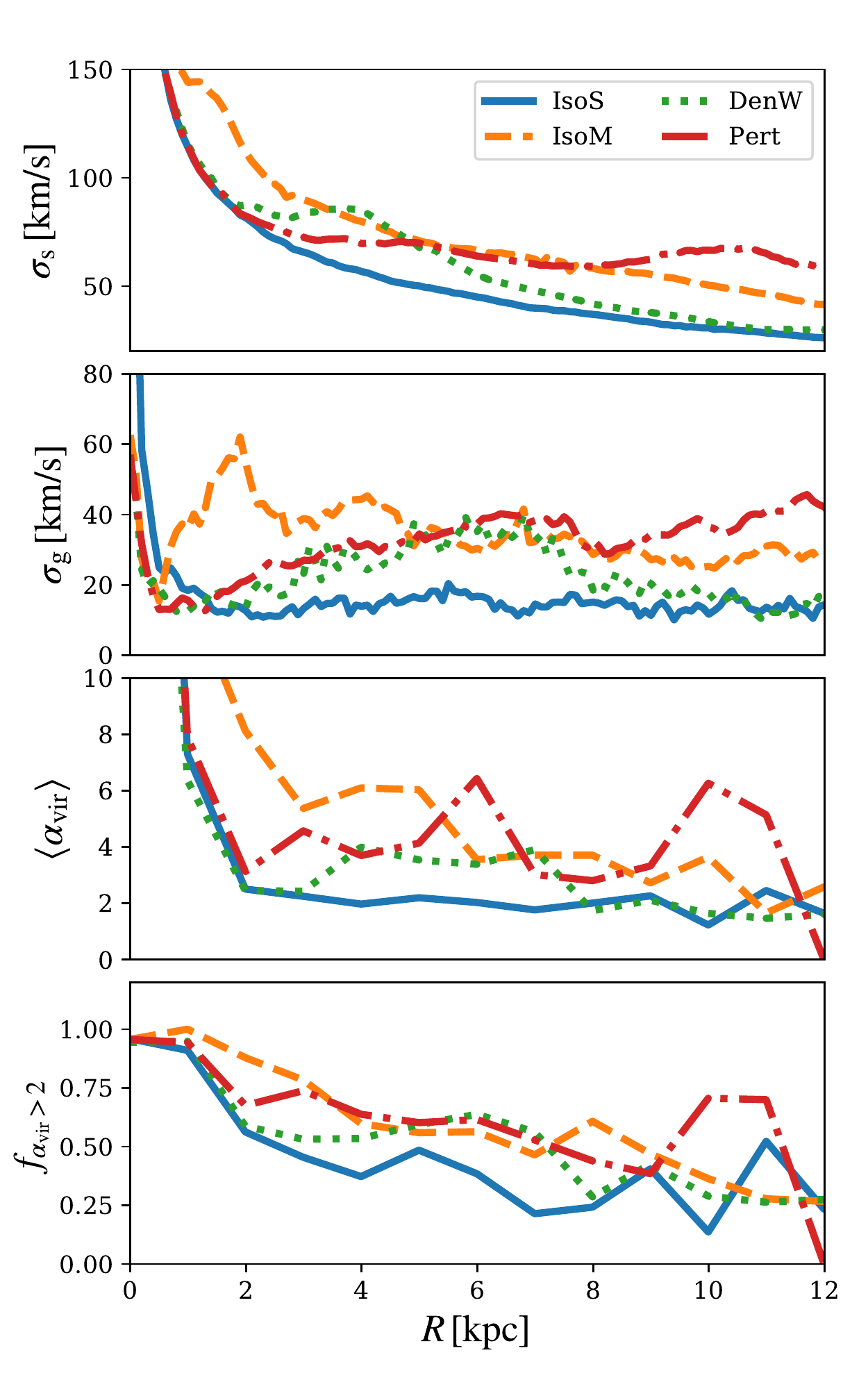}
\caption{From top: stellar velocity dispersion, gas velcoity dispersion, mean cloud virial parameter, and fraction of clouds with a virial parameter $>2$, as a function of galactic radius in each of the models.}
\label{dispR}
\end{centering}
\end{figure}

As the rotation curves are essentially the same in all simulations (i.e. identical epicycle frequency, $\kappa$; rotation frequency, $\Omega$; circular velocity, $V_c$), the velocity dispersion is one of the few remaining parameters that could be influencing cloud structure. Figure\;\ref{dispR} shows the 3D velocity dispersions as a function of radius ($\sigma_s$ and $\sigma_g$ in stars and gas respectively), plotted alongside the mean virial parameter ($\langle\alpha_{\rm vir}\rangle$) and fraction of clouds with a $\alpha_{\rm vir}> 2$ ($f_{\alpha_{\rm vir}>2}$). Figure\;\ref{starsigma} in Appendix\;\ref{AppA} shows maps of the separate radial and azimuthal stellar velocity dispersion for reference.

Each of the models show an interesting departure from $\sigma_s$ seen in IsoS. In IsoM the increase in velocity dispersion is everywhere, a response to the increased mass of the stellar disc. In Pert the increase is strongest in the outer disc, where the perturber has had the greatest impact. The DenW model only sees a significant increase in the radial range of $2\,{\rm kpc}<R<6\,{\rm kpc}$, where the spiral response is strongest, and also corresponds to the ILR (at 4\,kpc). The link between stellar velocity dispersion and spiral features has been documented in the literature \citep{1984ApJ...282...61S,2011ApJ...730..109F,2015MNRAS.453.1867G}.

The parameters $\sigma_g$, $\langle \alpha_{\rm vir}\rangle$, and $f_{\alpha_{\rm vir}>2}$ follow similar trends in each model, with the gas dispersion on galactic scales varying in the same manner as the internal dispersion of the clouds and the fraction of unbound clouds. For DenW the aforementioned peak in $\sigma_s$ is not coincident with where most of the unbound clouds lie, instead lying closer to 6\,kpc and the 4:1 resonance. 

There are some clear features between these models that tell the spiral mechanism apart. If there are clear peaks at some well defined radius then there is some clear resonance heating \citep{1972MNRAS.157....1L}, corresponding to a rigidly rotating spiral. If instead the dispersion rises to a plateau at the disc edge then a tidal response is triggering the formation of arms. If there is no such plateau or well defined peaks in $\sigma_{\rm s}$, instead steadily decreasing with radius in accordance with the stellar surface density, then the arms can be attributed to dynamically rotating features with no well defined pattern speed. This can also be interpreted as heating due to the superposition of a number of smaller transient waves, each with different co-rotation and resonance radii that span the whole disc, resulting in widespread disc heating \citep{2002MNRAS.336..785S}. Of course there is a caveat here that this is all simply a result of any kind of spiral pattern, with peaks in $\sigma_{\rm s}$ simply tracing where the spiral response is strongest. {While $\sigma_{\rm g}$ appears similar to $\sigma_{\rm s}$ it is significantly noisier, though difficulties with the observational determination of $\sigma_{\rm s}$ undermine its potential as a metric of the nature of the spiral arms.}
 
 It is not entirely clear to the authors why such a radial offset in the peaks in $\sigma_{\rm g}$ and $\sigma_{\rm s}$ exists in DenW, as the arms clearly dominate the disc at radii where both peak. As mentioned above, $\sigma_{\rm s}$ peaks at the ILR as expected. The outwards shift of the $\sigma_{\rm g}$ peak may be caused by feedback disrupting gas throughout the spiral arms, heating it up across a greater range in radii than what is driven by resonance heating in the $\sigma_{\rm s}$. To fully understand such features follow-up simulations are required with spirals of varying different pattern speeds, which we leave to a future work.

\subsection{Arm and interarm cloud spectra}
\label{sec_armspec}

\begin{figure}
\begin{centering}
\includegraphics[trim =10mm 0mm 10mm 0mm,width=41mm]{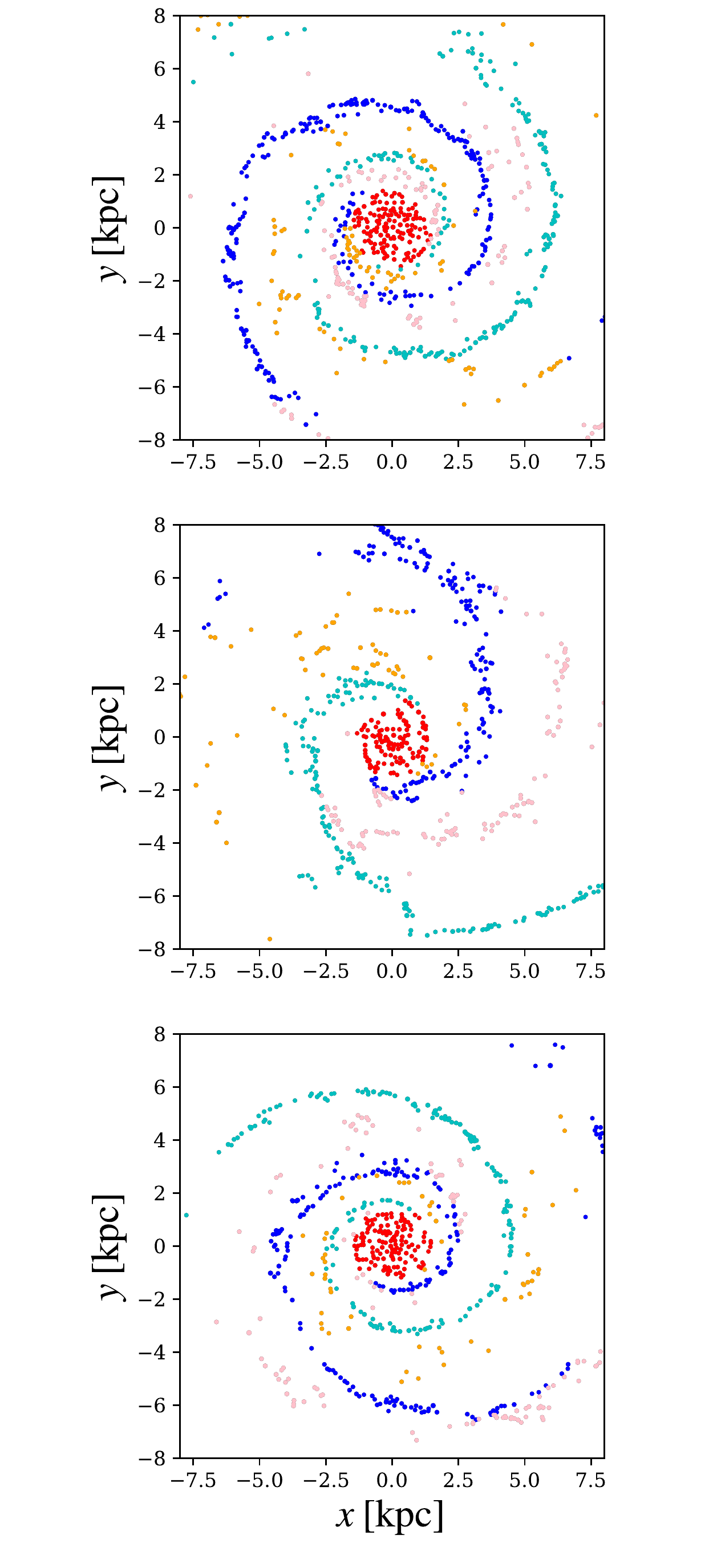}
\includegraphics[trim =10mm 0mm 10mm 0mm,width=41mm]{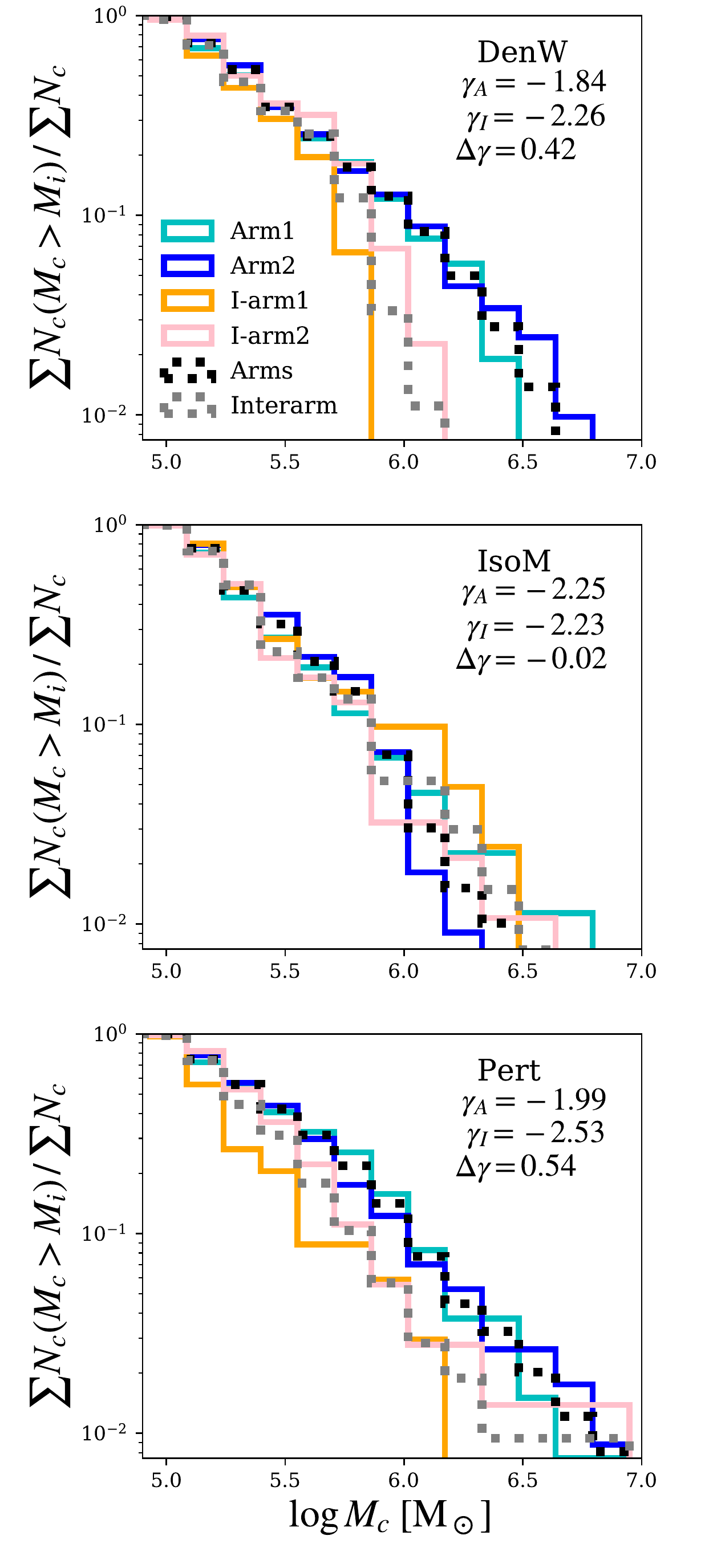}
\caption{Left: definitions of different arm regions. Right: normalised cumulative mass spectra for arms (cyan and blue) and interarm regions (orange and pink) for DenW, IsoM, and Pert models (top, middle and bottom respectfully). Grey and black dotted lines indicate the spectra for the combined interarm and arm populations respectively. Values in top-right corner indicate the fitted slope of mass functions of the arm and interarm regions and the difference between them: $\Delta\gamma=\gamma_A-\gamma_I$.}
\label{topgmc}
\end{centering}
\end{figure}

To further assess differences in cloud properties, we dissect the population into arm and interarm regions. Arms in the gas are defined via a Fourier decomposition technique, fitting to the $m=2$ component only, and only where that is the dominant mode (Table\;\ref{modelvalues}). This defines the gaseous arm, and we simply define the width as $\pm10$\arcdeg, a value that appears to encompass all arm structures across all radii in each model. This results in a linearly increasing physical arm width with radius, and values similar to those inferred for arms of the Milky Way \citep{2019ApJ...885..131R}. The arms are defined within $3\,{\rm kpc}\leq R\leq 8\,{\rm kpc}$ as all models display spiral features in this range.

In the left column of Figure\;\ref{topgmc} we show the top-down distribution of clouds defined by these criteria in each grand-design spiral model where clouds are colour-coded by region. The right column shows the normalised cumulative mass spectra for the two arms (blue and cyan), two interarm regions (pink and orange) and the combined arm and interam regions (black and grey dotted lines). The slopes in the mass spectra show slight differences between models, with values for the slopes given in the upper right for arm ($\gamma_A$) and interarm ($\gamma_I$) regions\footnote{We use the Nelder–Mead simplex algorithm in \textsc{python}’s \textsc{scipy} package and a $\chi^2$ minimizing statistic. Note this fit is to $\sum N_c(M_c>M_i)$, not the normalised spectra shown in Figure\;\ref{topgmc}.}. In particular the DenW and Pert models show arms having shallower slopes than the inter-arms ($\Delta \gamma=\gamma_A-\gamma_I\approx 0.5$), reaching up to higher cloud masses in the arms. IsoM by comparison has a similar slope for all four regions ($\Delta \gamma\approx 0$). While this could be attributed to the interarms including the tertiary interarm structure (pink) the slope for the other interarm region is also similar (yellow), which has no clear arm-like feature.

It appears the different spirals have different degrees of efficiency when it comes to building interarm clouds (IsoM$>$DenW$>$Pert), evident from the difference between interarm and arm spectra slopes $\Delta\gamma$ presented in Figure\;\ref{topgmc}. A simple link can be made to the duration gas is allowed to persist in the interarm regions. Consider first a diffuse ($\rho_c\sim1{\rm cm}^{-3}$) proto-cloud that is beginning to form upon entering the interarm region. The free-fall time of such a cloud is approximately $t_{\rm ff}=\sqrt{3\pi/32G\rho_c}\approx 120{\rm Myr}$. We can then consider how much time the cloud has before passing into an arm and a different environment. Approximating the arms as having widths of $\Delta \phi_{\rm arm}=20$\arcdeg{} and being two-fold symmetric in nature we calculate the interarm timeframe as:
\begin{equation}
t_{\rm IA}(R)= \frac{2\pi}{\Omega_{\rm g}(R)-\Omega_{\rm sp}(R)}
\frac{180^\circ-\Delta \phi_{\rm arm}}{180^\circ}
\end{equation}
where $\Omega_{\rm g}$ is the disc rotation frequency and $\Omega_{\rm sp}$ is the spiral pattern speed, which is different for each adopted spiral model. For DenW the arms have a rigid body rotation, with $t_{\rm IA}\rightarrow \infty$ moving out to co-rotation, which is just beyond the disc edge for these models. Moving to the centre $t_{\rm IA}$ drops to zero. For instance, $t_{\rm IA}({\rm 6\,kpc})\approx 135\,{\rm Myr}$. The arms in Pert rotate considerably slower outside of 4 kpc, rotating instead as $\Omega_{\rm g}-\kappa/2$, giving $t_{\rm IA}({\rm 6\,kpc})\approx 100\,{\rm Myr}$. IsoM by contrast has arms that rotate with $\Omega_{\rm sp}\approx \Omega_{\rm g}$ at all radii, so that the longevity of a spiral arm may be a suitable substitute metric of $t_{\rm IA}$, which is similar to the dynamical time in the disc: $t_{\rm IA}({\rm 6\,kpc})\sim 2\pi/{\Omega_{\rm g}}\approx 155\,{\rm Myr}$. Thus, our example diffuse proto-cloud would have plenty of time to collapse to higher densities in IsoM, marginally enough time in DenW and insufficient time in Pert. This agrees with what is seen in the slopes of the interarm mass spectra. Clouds in the interarms of IsoM reach similarly high masses as in the arms, the interarms of DenW build a decent population of low mass clouds but not as many high mass clouds as the arms, and the interam regions of Pert are less efficient at creating clouds of all masses. \citet{2015ApJ...806...72M} also use the inter-arm travel time of clouds to model the lifetimes of clouds, though focus instead on the radial dependence on different destruction processes in a single galaxy (M51), such as the effect of shear. The grand design spirals in this study have very similar values of shear in the interarm, and similarly low values in the arms. As such, it is unlikely shear is playing a role in any differences we see in cloud properties between the different models.

\subsection{GMCs around the spiral arms}
\label{sec_arm2}

\begin{figure*}
\begin{centering}
\includegraphics[trim =0mm 0mm 0mm 5mm,width=160mm]{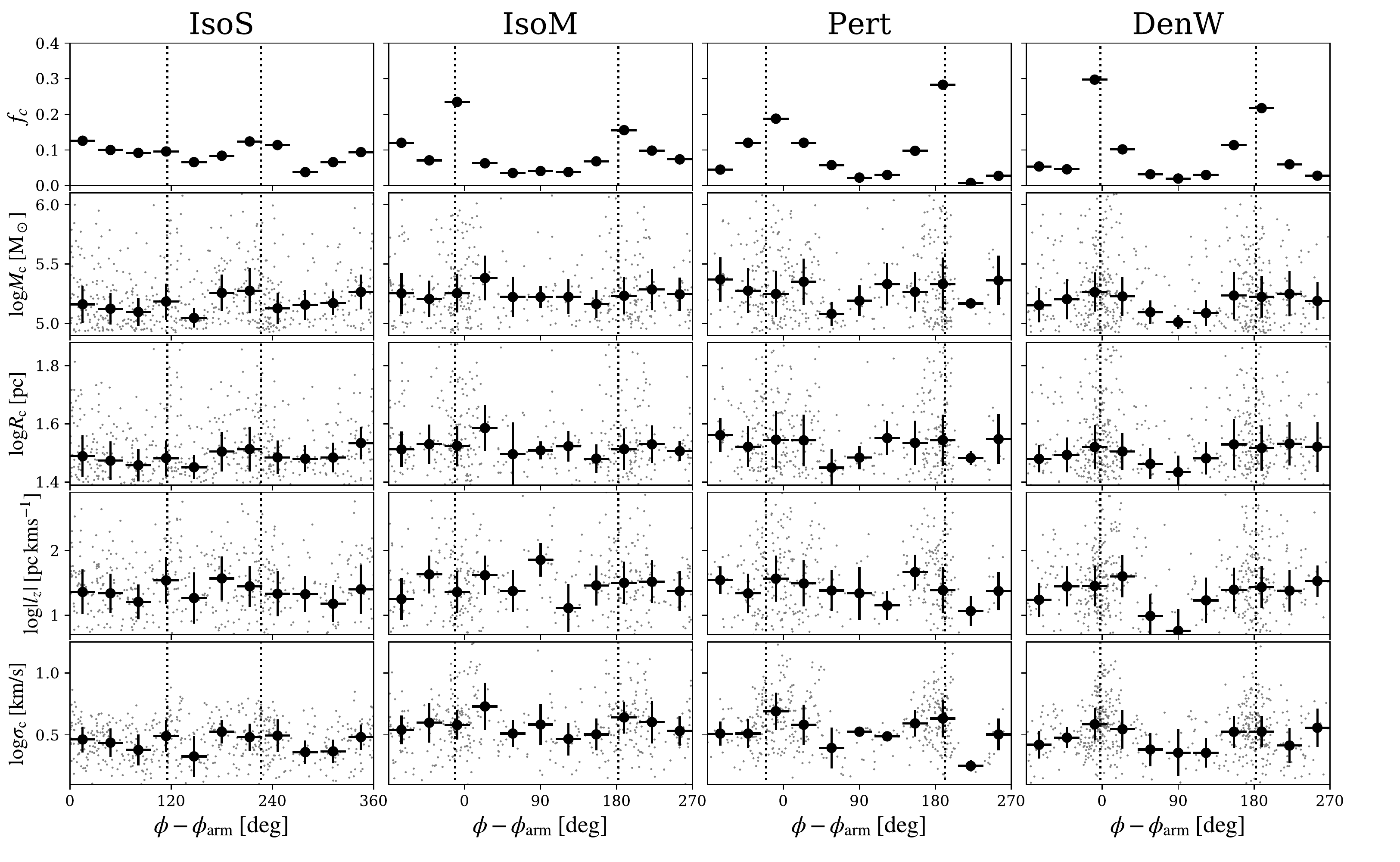}
\caption{Median properties of clouds binned over 2--8\,kpc against arm phase-shifted azimuth. Horizontal error bars indicate bin sizes and vertical are the median absolute deviations. Smaller grey points indicate the locations of the individual clouds. The top row shows the fraction of clouds. Angles are measured anti-clockwise, so gas moves from right to left in these panels. Dotted vertical lines show the azimuthal locations of the fitted logarithmic spirals, corresponding to the potential minima.}
\label{CloudsPhi}
\end{centering}
\end{figure*}

Figure\;\ref{CloudsPhi} shows various properties of clouds as a function of galactic azimuth, phase-shifted to the locations of the fitted spiral arms. The small grey points show the individual clouds, and the black points with error bars shown the median values over 12 azimuthal bins with vertical error bars indicating the median absolute deviation. The panels show, in order from top: $f_{\rm c}$ (the fraction of clouds at a given azimuth), $M_{\rm c}$, $R_{\rm c}$, $|l_{z}|$ and $\sigma_{\rm c}$.

Each grand design model shows clear and similar increases in $f_{\rm c}$ at the location of the spiral arms (vertical dotted lines). For each of the cloud properties there is a weak association with maximal values coinciding with spiral arms, and minima equidistant between the arms. This sinusoidal-like trend in properties is clearest in DenW, likely due to the very regular nature of the spiral pattern compared to Pert and IsoM, whose arm structures are more time-dependent. IsoM tends to show the weakest arm-interarm differences, in particular lacking the drop in properties around 180\arcdeg{} seen in Pert and DenW. The differing pattern speeds of the arms could be the cause of this, with gas accumulating into large complexes primarily as it passes into DenW and Pert arms. In IsoM the clouds in the interam regions have the autonomy to form massive clouds before the local material becomes part of a newly formed spiral arm. It may be expected that the streaming of gas into arms in DenW and Pert (and at differing rates between each) would manifest as a greater occurrence of cloud-cloud collisions, possibly evident in changes in $l_z$. However, we see no convincing evidence for this, concluding that the nature of spiral arms does little to impact the changes in individual cloud properties between arm and interarm regions.

Ultimately we suffer from quite small statistics in the inter-arm regions due to our gas resolution. This would benefit from a higher resolution follow-up specifically targeting the capacity of inter-arm regions to host GMCs.

\subsection{Cloud rotation}
\label{sec_rot}

\begin{figure*}
\begin{centering}
\includegraphics[trim =10mm 0mm 10mm 0mm,width=120mm]{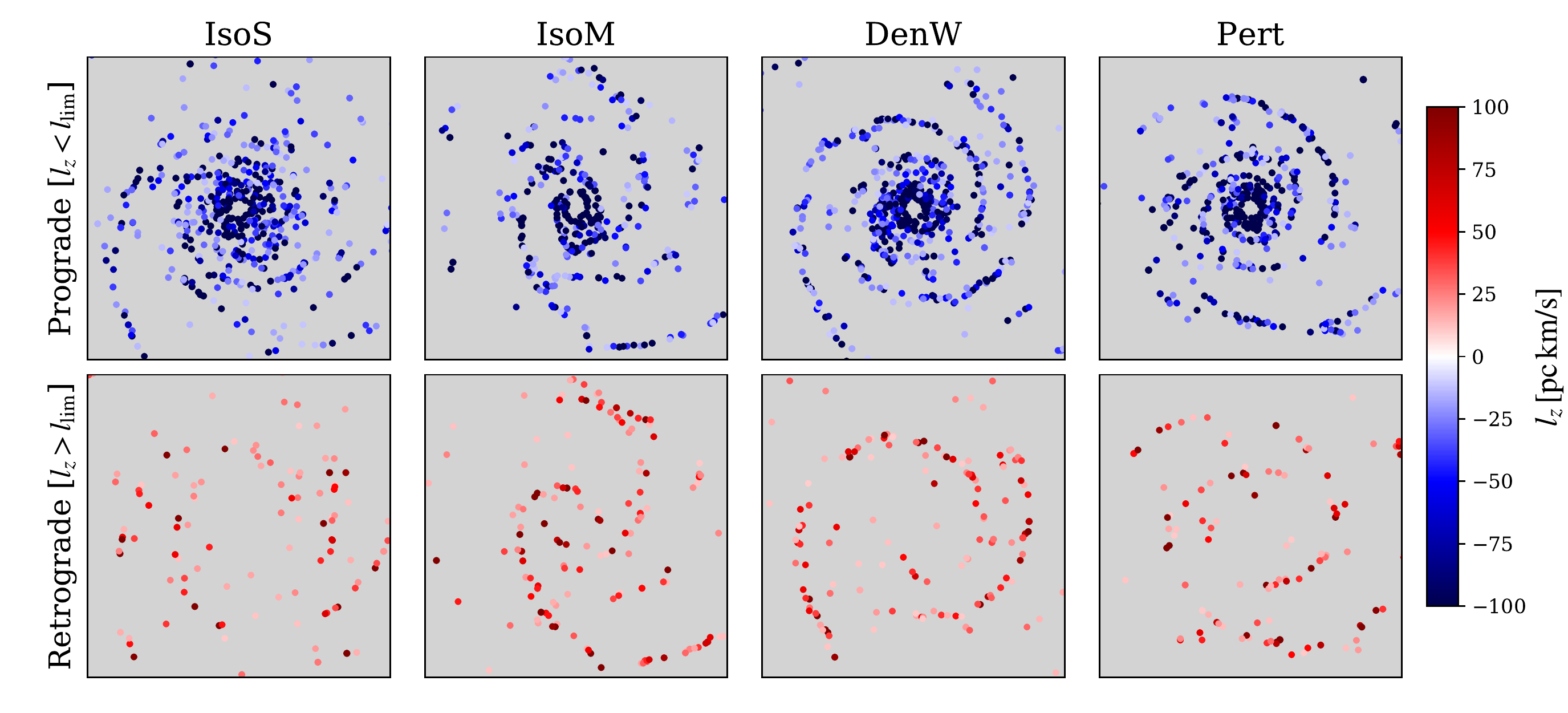}
\caption{Distribution of prograde (top) and retrograde (bottom) rotating clouds in each simulation. A cut of $l_{\rm lim}=\pm 10{\rm pc\, km/s}$ is used to remove population with the smallest rotations.}
\label{proretTop}
\end{centering}
\end{figure*}

\begin{figure}
\begin{centering}
\includegraphics[trim =10mm 0mm 10mm 0mm,width=70mm]{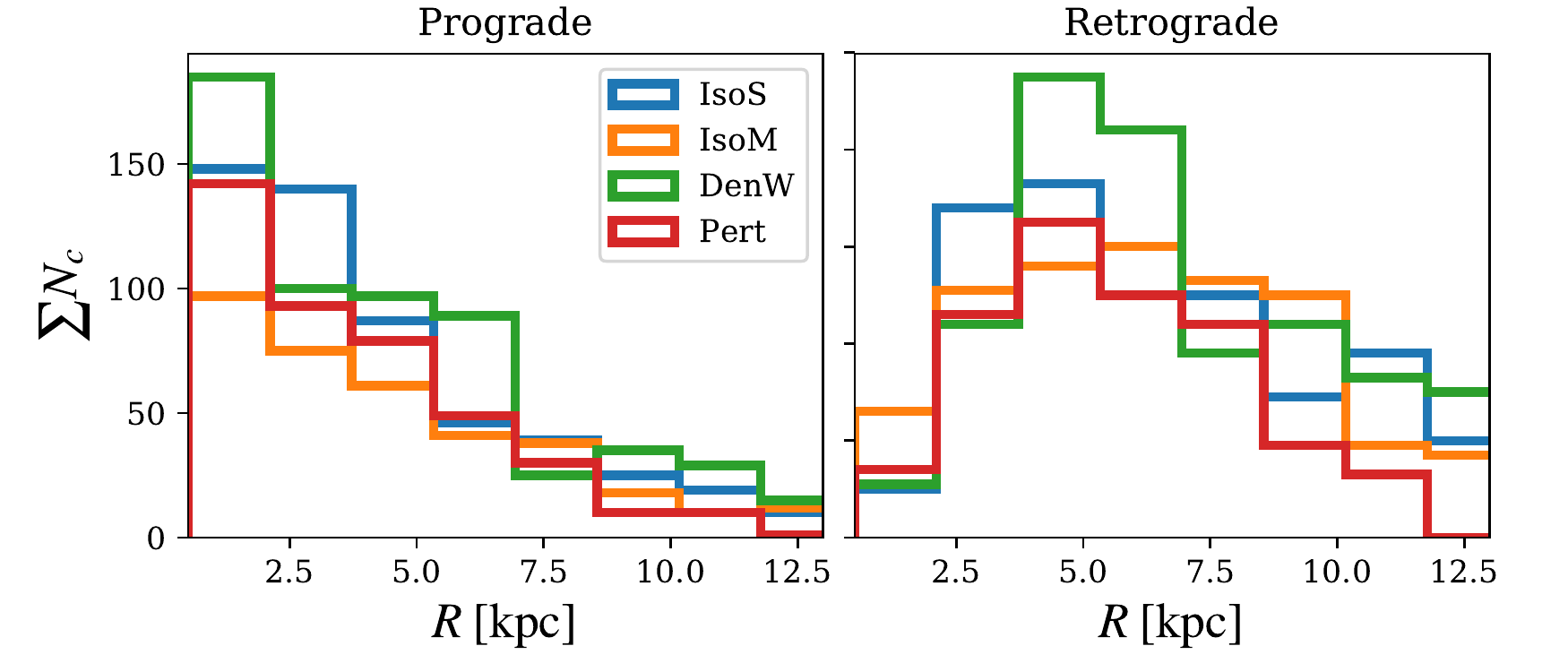}
\includegraphics[trim =10mm 0mm 10mm 0mm,width=85mm]{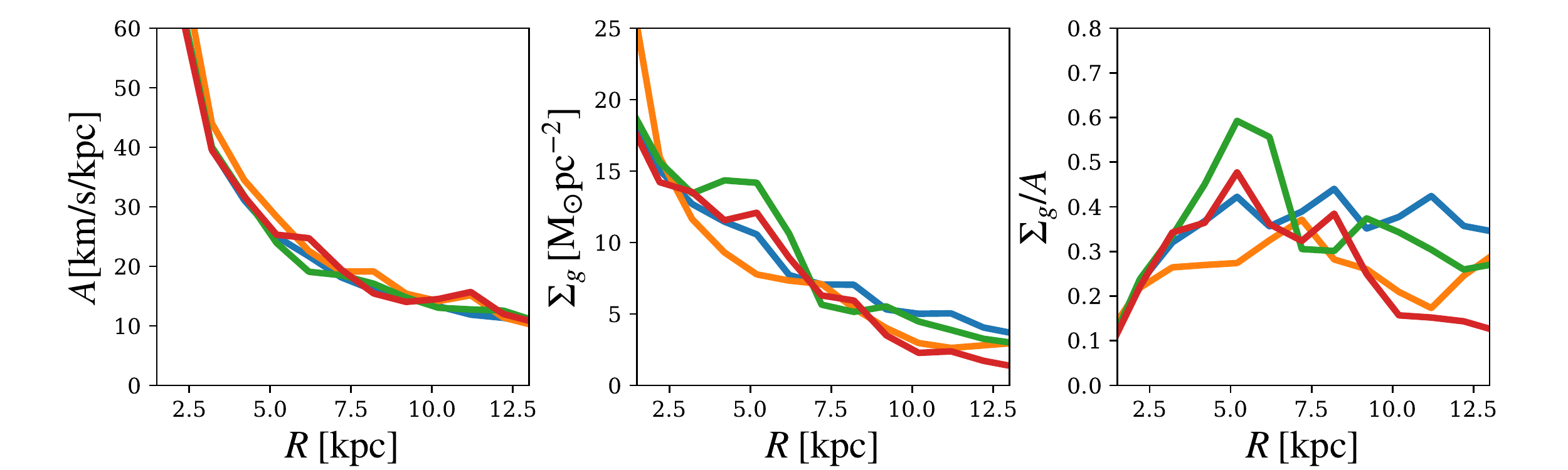}
\caption{Top: radial distribution of prograde rotating clouds (left) and retrograde rotating clouds (right) for each of the four models. Bottom: radial dependence of shear (Oort's $A$ constant), gas surface density and their ratio.}
\label{proret}
\end{centering}
\end{figure}

\begin{table}
\centering
 \begin{tabular}{@{}l |c  c c }
  Model & Whole disc & Arms & Interarm \\
  \hline
  IsoM & 0.38 & 0.41 & 0.39 \\
  DenW & 0.28 & 0.33 & 0.30 \\
  Pert & 0.26 & 0.34 & 0.27
 \end{tabular}
 \caption{Fraction of retrograde clouds in each environment in the grand design spiral models. The total amount includes the inner disc, whereas arm and inter-arm effectively dissect the outer disc into two regions.}
 \label{rottable}
\end{table}

Recently \citet{2020A&A...633A..17B} studied the rotational properties (i.e. the spin) of clouds in M51. They found that the rotation of clouds is tied to the spiral structure and the shear in the disc. Interestingly they also notice a trend in retrograde rotating clouds to be more common in spiral arm regions, with a roughly 10\% higher retrograde cloud fraction in the arms than interarms. Using our definition of arm locations, and summing over all radii, we find the fraction of retrograde clouds as given in Table\;\ref{rottable}. We do see a general trend of arm clouds having a higher fraction of retrograde rotation compared to the inter-arms, though the magnitude of the difference is somewhat smaller. Figure\;\ref{proretTop} explicitly shows the locations of prograde (top) and retrograde (bottom) rotating clouds from a top-down perspective. Clouds are coloured by angular momentum, with each column showing a different model. 

\begin{figure}
\begin{centering}
\includegraphics[trim =10mm 0mm 10mm 0mm,width=85mm]{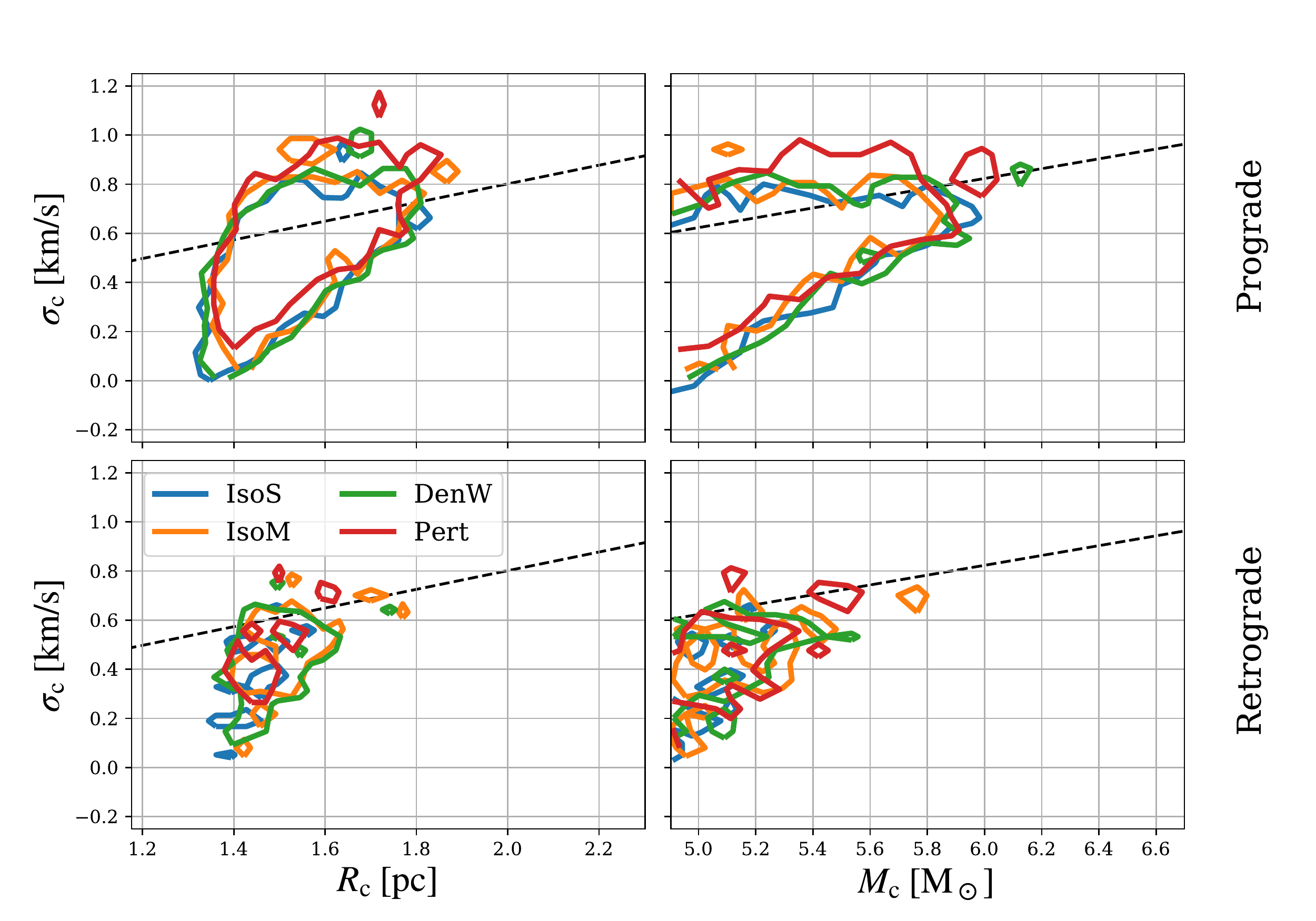}
\caption{Larson's scaling relations \citep{1981MNRAS.194..809L} for both retrograde and prograde cloud populations. The scaling relations $\sigma_{\rm c}=0.42 M_{\rm c}^{0.20}$ and $\sigma_{\rm c}=1.10R_{\rm c}^{0.38}$ are shown as dashed lines.}
\label{LarsonPR}
\end{centering}
\end{figure}

Retrograde clouds appear to be preferentially located in the mid/outer disc in all models ($R>4$\,kpc). Figure\;\ref{proret} shows clouds binned by radius in each model and separated by pro- and retrograde rotators. There is a clear trend in that the distribution of prograde rotating clouds effectively trace the gas surface density, regardless of the different arm perturbations. The retrograde clouds however are preferentially located in the mid-disc. In \citet{2018MNRAS.480.3356P} (Fig.\;9) we presented a map of the shear (i.e. Oort's constant: $A=-\frac{1}{2}R\frac{d\Omega}{d R}$) for the same rotation curve adopted here, which seems to anti-correlate with the radial trend seen here for the retrograde clouds. To illustrate this the lower panels show the radial distribution of $\Sigma_g$, $A$ and $\Sigma_g/A$. A very good correspondence between $\Sigma_{g}/A$ and the distribution of retrograde clouds is seen (as well as the prograde distribution with $\Sigma_g$), with IsoM and IsoS showing flatter distributions and DenW and Pert both peaking around 5\,kpc and dropping faster in the outer disc. This indicates that shear rates of approximately $>40\,{\rm km/s/kpc}$ can prevent the generation of a significant retrograde cloud population. The simple explanation is that shear would shred apart any clouds that are rotating in the opposite sense to the disc, with passing gas and stars inducing torques that oppose their retrograde spin. In the outer disc the shear is lower, negating this effect. In addition, the simulations of \citet{2015MNRAS.446.3608D} showed that cloud collisions/mergers are more frequent in armed than arm-free gas discs, which could be a progenitor of retrograde clouds. As spiral arms tend to dominate the mid/outer regions of our models, this could also be why retrograde clouds are preferentially located away from the inner disc. It is unlikely that an over-crowding effect in spiral arms alone is the culprit for the retrograde overdensity around 5\,kpc, as it is present in the IsoS model (which displays only weak spiral arms) just as clearly as in the grand design models.

To a first approximation the rotation curves here are effectively flat, which result in the relations $A\approx\Omega_{\rm g} /2$ and $\kappa\approx\sqrt 2 /\Omega_{\rm g}$, which makes Toomre's stability criterion in the gas:
\begin{equation}
    Q_{\rm g} = \frac{  c_{\rm eff} \kappa   }{     \pi G \Sigma_{\rm g}  } \approx 
  \frac{   2 \sqrt{2}A c_{\rm eff}  }{     \pi G \Sigma_{\rm g}  }
\end{equation}
where $c_{\rm eff}$ is the effective sound speed of the gas. This implies $\Sigma_{\rm g}/A \propto c_{\rm eff}/Q_{\rm g}$, which is seen to be effectively the case in our simulated discs, with slight differences in the outer regions where the rotation curve begins to decline. This relation suggests that higher non-circular motions implied by higher $c_{\rm eff}(\sigma_{\rm g})$ increases the population of retrograde clouds, for galaxies with equivalent values for $Q_{\rm g}$.

Inspection of the properties of the retrograde clouds (e.g. Larson's scaling relations, as shown in Figure\;\ref{LarsonPR}) indicates they are simply sampling the same distributions as the prograde clouds. They are not preferentially more massive or larger than the prograde sample in any of the four models. They are also not seen to be in any particular location azimuthally with respect to the spiral arms themselves.

\section{Discussion}
\label{sec:disc}

\begin{table*}
\def\arraystretch{1.5}
\centering
 \begin{tabular}{@{}c | c | c c c}
 
  \multirow{2}{*}{\bf Nature of arm} & \multirow{2}{*}{\bf  Dynamics} & \multicolumn{3}{c}{\bf  Observational diagnostics}  \\
           &             &  Disc stars & ISM & GMCs  \\
  
  \hline
  Dynamic 
  & \makecell{Large-scale spiral \\ displays material rotation, \\ lifetime: $\mathcal{O}(100\,{\rm Myr})$} 
  & \makecell{Little heating of disc,\\ global radial migration}
  & \makecell{No spurs,\\ weak shocks,\\ no systematic star/gas offset} 
  & \makecell{Interarm and arm \\ mass spectra similar,\\ weak peaks in mass and \\ virial parameter in arms} 
  \\
  &&&&\\
  \makecell{Density \\ Wave} 
  & \makecell{$\Omega_{\rm p}$ constant/\\ nearly constant,\\ lifetime: $\mathcal{O}({\rm Gyr})$} 
  & \makecell{Heating at \\ I/OLR}
  & \makecell{Strong shocks, \\ clear spur features,\\ clear star/gas offset} 
  & \makecell{Less favourable to forming \\ highest mass clouds,\\ clear arm/interarm change\\ in cloud properties}
  \\
  &&&&\\
  Tidal 
  & \makecell{$\Omega_{\rm p}$ decreases with radius as \\ classical kinematic wave, \\ lifetime: $\mathcal{O}(\lesssim 1\,{\rm Gyr})$} 
  & \makecell{Heating in \\ outer disc}
  & \makecell{Shocks, weak spurs,\\ asymmetric arm response \\in the bridge/tail}
  & \makecell{Reduced population of \\small and well \\ bound clouds}
  \\
 \end{tabular}
 \caption{Properties and possible observable signatures of the spirals models discussed in this work, including numerous inferences from the studies in the literature. Properties are focussed on grand design, 2-armed spirals.}
 \label{whatspiral}
\end{table*}

Table\;\ref{whatspiral} summarises more broadly the results presented here and in other studies in relation to different types of spiral arms. We stress this table does not list findings solely reported in this work, but rather draws together a number properties of different kinds of spiral arms which are present in the simulations shown here (if not explicitly discussed). The table is also not exhaustive, focusing more on properties that directly tie into this study in some way. These focus on the properties of 2-armed spiral patterns (tidal interactions are unlikely to have a direct bearing on generating 4-armed spiral patterns), though the listed properties are not inherently tied to the number of arms exhibited. We do not list arm generation mechanisms not studied in this work, such as those driven by bars \citep{1976ApJ...209...53S}, triaxial dark matter halos \citep{2012ARep...56...16K} or as a response to disc clumping \citep{2013ApJ...766...34D}.

It is most likely the case that any one of these mechanisms is not responsible for all observed spiral arms seen in nature. M51 in particular is a prime example of this. This galaxy exhibits differing arm and interarm mass spectra \citep{2014ApJ...784....3C}, as we report here in Sec.\;\ref{sec_armspec}. It also shows different features in the ISM of each arm \citep{2017MNRAS.465..460E} as well as spurring features in the arms \citep{2017ApJ...836...62S}, both similar features to those seen in the numerical work of \citet{2017MNRAS.468.4189P}.

Studies of the Milky Way's ISM and GMC population are hampered by the Earth's location within the disc, with observational studies offering conflicting findings regarding differences between the arm and interarm regions \citep{2019MNRAS.483.4291C,2019A&A...632A..58R}. There is a somewhat limited quantity of high quality (i.e. cloud-scale) ISM observational data for unbarred grand design spirals, with M51 the notable exception. However, upcoming results from the PHANGS-ALMA survey includes promising targets such as M74 and NGC\,4254, and several galaxies in their survey show spatial offsets between ISM and star formation traces which also acts as a metric for density wave-style arms (\citealt{2019ApJ...887...49S}, see also \citealt{2016ApJ...827L...2P}). High resolution ISM observations presented in \citet{2016ApJ...831...16L} hint at differences in the velocity dispersion and virial parameter at cloud scales in the spirals M74, M51 and M33. M74 has the most well bound ISM (their figure 10), followed by M51 and M33. Comparison of virial parameters in this study would imply M74 has more density-wave like spiral arms, and higher virial parameters in M51 and M33 would imply a tidal or dynamic kind of arm. The existence of clear spurs would then point to M51 as tidally driven and M33 as being driven by disc instabilities (the latter was also inferred by \citealt{2018MNRAS.478.3793D}, though stellar feedback also played a large role in sculpting arms in their simulations). Pattern speeds and arm winding rates offer some of the most promising direct evidence of the origin of a spiral arm, and there seems to be mixed evidence regarding whether arms rotate at a constant pattern speed \citep{2019MNRAS.487.1808M,2019MNRAS.490.1470P,2019NatAs...3..178P}. The resolution of modern facilities such as ALMA appears sufficient to resolve the ISM and GMC diagnostics in Table\;\ref{whatspiral}, making a larger galaxy sample size the important next step in testing these criteria.

Using stellar material to tell spiral mechanisms apart is also a possibility, though the necessary observational data is even harder to come by than that of the ISM. The Gaia data for $\sigma_{v_r}$ presented in 
\citet{2018A&A...616A..11G} shows a clear peak in an arm-like feature just inside the Solar radius (and to a lesser extent in $\sigma_{v_\phi}$). If these peaks are associated with the resonances of a density wave-like spiral pattern (Figures\;\ref{dispR} and \ref{starsigma}) then they would correspond to a pattern speed for a 2-armed spiral that is much lower than suggested by contemporary studies. For instance, using the rotation curve of \citet{2019ApJ...871..120E} gives an ILR of 2.8\,kpc for a pattern speed of 20\ps{}. However, a 4-armed spiral corresponds to a ILR (now the 4:1 resonance) at 7.5\,kpc, which is extremely close to the peaks in velocity dispersion seen in Gaia DR2. Ultimately Milky Way diagnostics will always be plagued by our indirect measurements of its morphology, making external systems more promising targets. Data from large surveys such as SDSS-MANGA \citep{2015ApJ...798....7B} are another avenue for probing the relation between stellar velocity dispersions and spiral structure, though ultimately still lack the resolution to detect spiral arm resonance heating. The radial dependance of the stellar velocity dispersion in 34 CALIFA galaxies in presented in \citet{2019MNRAS.489.3797M}, but dispersion values are limited to a resolution of $\approx70\,{\rm km\,s^{-1}}$, whereas the signatures in the models presented in this work are of the order of a few 10s of ${\rm km\,s^{-1}}$.  Age/colour gradients in stars are also possible diagnostics of spiral features, with measurements also not pointing towards a single mechanism operating across all galaxies \citep{2009ApJ...694..512M,2018MNRAS.478.3590S}. 

There are clear avenues for follow-up studies to build upon the results presented here. Moving to higher resolution is an obvious area of improvement, possibly focussing on high resolution patches of the arms in each model, assessing the time-evolution of clouds as a result of different arm environments. A more consistent comparison to observational data products is also necessary for a more complete picture, both on cloud \citep{2015MNRAS.453.3082P,2016MNRAS.458.3667D} and entire galactic scales \citep{2010MNRAS.406.1460A}, including the use of more comparable tracers such as CO or H$_2$ column densities. Such works will implement improvements to the somewhat rudimentary subgrid physics used here, such as the addition of a consistent treatment of photo-electric heating \citep{2019arXiv191105252B}, parameter free supernova heating \citep{2014MNRAS.442.3013K}, and early stellar/\ion{H}{ii}-region feedback \citep{2013MNRAS.428..129S}. While these self-consistent galactic models are enlightening, the nature of the stellar disc response and discretised $N$-body halo/disc make it difficult to control certain aspects of the arms like pitch angle and strength. A more controlled study is underway where a large variety of different spiral perturbations are represented instead via analytic models, rather than depending on the somewhat uncontrollable response of the stellar and dark matter distribution (expanding to include arms made by bars and triaxial dark halos), allowing for a wider-scope comparison than presented here.

\section{Conclusions}
\label{SecConc}

We have analysed simulations investigating the formation of giant molecular clouds and structure of the interstellar medium in galaxies with different spiral arm formation mechanisms, but whose properties otherwise match as closely as possible. These include arms formed by a rigidly rotating density wave, induced by a passing companion, and formed by instabilities caused by the self-gravity of the unperturbed stellar disc itself. Our key findings are as follows:

\begin{enumerate}
\item Grand design spiral arms tend to allow the formation of more massive and poorly bound GMCs (Figure\;\ref{ModelHistCompV}, \ref{alphamap}). The cloud population is seemingly robust, to first order, when exposed to a strong spiral perturbation. Distributions of cloud properties are relatively unchanged between different spiral arm models, with the largest change seen between the flocculent disc and the three grand design spiral models. This is also true for the structure of the ISM and its kinematics in general (Figures\;\ref{denmap}, \ref{gasstream} and \ref{AzVels}), making each model hard to distinguish.

\item Some secondary differences are seen in the tails of the distributions, with the tidally perturbed disc generally showing a stronger preference for hosting massive and weakly bound clouds compared to the other models (Figure\;\ref{ModelHistCompV}). The tidal and dynamic arms seem deficient in their low mass cloud population compared to the flocculent disc, while the density wave arms can maintain their low mass population while simultaneously pushing into higher masses.

\item Each spiral model drives differences in the kinematic heating of the gas and stars. The tidally perturbed arms tend to heat the outer disc, while a steadily rotating density wave heats primarily at resonance locations (Figure\;\ref{dispR}). Such features also correlate with the location of the more loosely bound GMCs.

\item The cloud mass spectra for each model is quite similar, but shows differences in the interarm regions (Figure\;\ref{topgmc}). The tidal spiral has the steepest interarm spectra, and the dynamic spiral has interarm spectra nearly identical to that of the arms. This can be attributed to how long clouds have to form, which differs in each model due to the differing spiral pattern speeds with radius. The dynamic-style arms have the weakest contrast between arm and interarm cloud properties in general (Figure\;\ref{CloudsPhi}).

\item We see almost all retro-grade clouds confined to the mid/outer disc in all models, implying strong inner shear in the inner disc acts to prevent the formation of retrograde clouds (Figure\;\ref{proret}). Retrograde clouds are also preferentially situated in the arms themselves (Figure\;\ref{proretTop}), with a distribution that closely follows the ratio of surface density to shear. This implies only spiral arms have a high enough gas concentration to mitigate the destructive shear that inhibits retrograde cloud formation that is presumably triggered by cloud collisions and mergers. 

\end{enumerate}

We have newly presented, in particular, two different metrics of spiral structure: the variations in velocity dispersion and the slope of the inter-arm cloud mass spectra. These metrics, and those listed in Table\;\ref{whatspiral}, could have great utility in identifying the nature of spiral structure in modern surveys. Unearthing any kind of diagnostic of spiral structure theory is of prime importance in this era of high resolution Milky Way and galactic-observations.

\section*{Acknowledgments}

We thank the anonymous referee for their reading of this manuscript and insightful comments/suggestions that have improved this work.
Special thanks to the authors of \textsc{gasoline2} \citep{2017MNRAS.471.2357W} for making this study possible. We utilised the \textsc{pynbody} package \citep{2013ascl.soft05002P} for post-processing and analysing of the \textsc{tipsy} files created by \textsc{gasoline2}. Numerical computations were carried out on Cray XC50 at Center for Computational Astrophysics, National Astronomical Observatory of Japan.

ARP acknowledges the support of The Japanese Society for the Promotion of Science (JSPS) KAKENHI grant for Early Career Scientists (20K14456). CLD acknowledges funding from the European Research Council for the Horizon 2020 ERC consolidator grant project ICYBOB, grant number 818940. JB acknowledges the support by JSPS KAKENHI Grant Number 18H01248. DC acknowledges the support by the German \emph{Deut\-sche For\-schungs\-ge\-mein\-schaft, DFG\/} project number SFB956A. ADC acknowledges the support from the Royal Society University Research Fellowship (URF/R1/191609). AH is funded by the JSPS KAKENHI Grant (JP19K03923). FE is supported by JSPS KAKENHI Grant Number 17K14259. This project was developed in part at the 2019 Santa Barbara Gaia Sprint, hosted by the Kavli Institute for Theoretical Physics at the University of California, Santa Barbara. This research was supported in part by the Heising-Simons Foundation and the National Science Foundation under grant no. NSF PHY-1748958. We thank Yusuke Fujimoto and Elizabeth Tasker for useful discussions and comments related to this work.

\section*{Data availability}
The data underlying this article will be shared on reasonable request to the corresponding author.

\bibliographystyle{mnras}
\bibliography{Pettitt_gmc-models.bbl}

\begin{thebibliography}{}
\makeatletter
\relax
\def\mn@urlcharsother{\let\do\@makeother \do\$\do\&\do\#\do\^\do\_\do\%\do\~}
\def\mn@doi{\begingroup\mn@urlcharsother \@ifnextchar [ {\mn@doi@}
  {\mn@doi@[]}}
\def\mn@doi@[#1]#2{\def\@tempa{#1}\ifx\@tempa\@empty \href
  {http://dx.doi.org/#2} {doi:#2}\else \href {http://dx.doi.org/#2} {#1}\fi
  \endgroup}
\def\mn@eprint#1#2{\mn@eprint@#1:#2::\@nil}
\def\mn@eprint@arXiv#1{\href {http://arxiv.org/abs/#1} {{\tt arXiv:#1}}}
\def\mn@eprint@dblp#1{\href {http://dblp.uni-trier.de/rec/bibtex/#1.xml}
  {dblp:#1}}
\def\mn@eprint@#1:#2:#3:#4\@nil{\def\@tempa {#1}\def\@tempb {#2}\def\@tempc
  {#3}\ifx \@tempc \@empty \let \@tempc \@tempb \let \@tempb \@tempa \fi \ifx
  \@tempb \@empty \def\@tempb {arXiv}\fi \@ifundefined
  {mn@eprint@\@tempb}{\@tempb:\@tempc}{\expandafter \expandafter \csname
  mn@eprint@\@tempb\endcsname \expandafter{\@tempc}}}

\bibitem[\protect\citeauthoryear{{Acreman}, {Douglas}, {Dobbs}  \&
  {Brunt}}{{Acreman} et~al.}{2010}]{2010MNRAS.406.1460A}
{Acreman} D.~M.,  {Douglas} K.~A.,  {Dobbs} C.~L.,   {Brunt} C.~M.,  2010,
  \mn@doi [\mnras] {10.1111/j.1365-2966.2010.16858.x}, \href
  {http://adsabs.harvard.edu/abs/2010MNRAS.406.1460A} {406, 1460}

\bibitem[\protect\citeauthoryear{{Baba}, {Saitoh}  \& {Wada}}{{Baba}
  et~al.}{2013}]{2013ApJ...763...46B}
{Baba} J.,  {Saitoh} T.~R.,   {Wada} K.,  2013, \mn@doi [\apj]
  {10.1088/0004-637X/763/1/46}, \href
  {http://adsabs.harvard.edu/abs/2013ApJ...763...46B} {763, 46}

\bibitem[\protect\citeauthoryear{{Baba}, {Morokuma-Matsui}  \& {Egusa}}{{Baba}
  et~al.}{2015}]{2015PASJ...67L...4B}
{Baba} J.,  {Morokuma-Matsui} K.,   {Egusa} F.,  2015, \mn@doi [\pasj]
  {10.1093/pasj/psv048}, \href
  {http://adsabs.harvard.edu/abs/2015PASJ...67L...4B} {67, L4}

\bibitem[\protect\citeauthoryear{{Baba}, {Morokuma-Matsui}, {Miyamoto}, {Egusa}
   \& {Kuno}}{{Baba} et~al.}{2016}]{2016MNRAS.460.2472B}
{Baba} J.,  {Morokuma-Matsui} K.,  {Miyamoto} Y.,  {Egusa} F.,   {Kuno} N.,
  2016, \mn@doi [\mnras] {10.1093/mnras/stw987}, \href
  {http://adsabs.harvard.edu/abs/2016MNRAS.460.2472B} {460, 2472}

\bibitem[\protect\citeauthoryear{{Baba}, {Morokuma-Matsui}  \& {Saitoh}}{{Baba}
  et~al.}{2017}]{2017MNRAS.464..246B}
{Baba} J.,  {Morokuma-Matsui} K.,   {Saitoh} T.~R.,  2017, \mn@doi [\mnras]
  {10.1093/mnras/stw2378}, \href
  {http://adsabs.harvard.edu/abs/2017MNRAS.464..246B} {464, 246}

\bibitem[\protect\citeauthoryear{{Baba}, {Kawata}, {Matsunaga}, {Grand }  \&
  {Hunt}}{{Baba} et~al.}{2018}]{2018ApJ...853L..23B}
{Baba} J.,  {Kawata} D.,  {Matsunaga} N.,  {Grand } R. J.~J.,   {Hunt} J.
  A.~S.,  2018, \mn@doi [\apjl] {10.3847/2041-8213/aaa839}, \href
  {https://ui.adsabs.harvard.edu/abs/2018ApJ...853L..23B} {853, L23}

\bibitem[\protect\citeauthoryear{{Benincasa} et~al.,}{{Benincasa}
  et~al.}{2019a}]{2019arXiv191105251B}
{Benincasa} S.~M.,  et~al., 2019a, arXiv e-prints, \href
  {https://ui.adsabs.harvard.edu/abs/2019arXiv191105251B} {p. arXiv:1911.05251}

\bibitem[\protect\citeauthoryear{{Benincasa}, {Wadsley}, {Couchman}, {Pettitt},
  {Keller}, {Woods}  \& {Grond}}{{Benincasa}
  et~al.}{2019b}]{2019arXiv191105252B}
{Benincasa} S.~M.,  {Wadsley} J.~W.,  {Couchman} H. M.~P.,  {Pettitt} A.~R.,
  {Keller} B.~W.,  {Woods} R.~M.,   {Grond} J.~J.,  2019b, arXiv e-prints,
  \href {https://ui.adsabs.harvard.edu/abs/2019arXiv191105252B} {p.
  arXiv:1911.05252}

\bibitem[\protect\citeauthoryear{{Binney} \& {Tremaine}}{{Binney} \&
  {Tremaine}}{1987}]{1987gady.book.....B}
{Binney} J.,  {Tremaine} S.,  1987, {\emph{Galactic Dynamics}}.
{Princeton, NJ, Princeton University Press}

\bibitem[\protect\citeauthoryear{{Boily}, {Kroupa}  \&
  {Pe{\~n}arrubia-Garrido}}{{Boily} et~al.}{2001}]{2001NewA....6...27B}
{Boily} C.~M.,  {Kroupa} P.,   {Pe{\~n}arrubia-Garrido} J.,  2001, \mn@doi
  [\na] {10.1016/S1384-1076(01)00039-2}, \href
  {http://adsabs.harvard.edu/abs/2001NewA....6...27B} {6, 27}

\bibitem[\protect\citeauthoryear{{Braine}, {Hughes}, {Rosolowsky}, {Gratier},
  {Colombo}, {Meidt}  \& {Schinnerer}}{{Braine}
  et~al.}{2020}]{2020A&A...633A..17B}
{Braine} J.,  {Hughes} A.,  {Rosolowsky} E.,  {Gratier} P.,  {Colombo} D.,
  {Meidt} S.,   {Schinnerer} E.,  2020, \mn@doi [\aap]
  {10.1051/0004-6361/201834613}, \href
  {https://ui.adsabs.harvard.edu/abs/2020A&A...633A..17B} {633, A17}

\bibitem[\protect\citeauthoryear{{Bundy} et~al.,}{{Bundy}
  et~al.}{2015}]{2015ApJ...798....7B}
{Bundy} K.,  et~al., 2015, \mn@doi [\apj] {10.1088/0004-637X/798/1/7}, \href
  {https://ui.adsabs.harvard.edu/abs/2015ApJ...798....7B} {798, 7}

\bibitem[\protect\citeauthoryear{{Choi}, {Dalcanton}, {Williams}, {Weisz},
  {Skillman}, {Fouesneau}  \& {Dolphin}}{{Choi}
  et~al.}{2015}]{2015ApJ...810....9C}
{Choi} Y.,  {Dalcanton} J.~J.,  {Williams} B.~F.,  {Weisz} D.~R.,  {Skillman}
  E.~D.,  {Fouesneau} M.,   {Dolphin} A.~E.,  2015, \mn@doi [\apj]
  {10.1088/0004-637X/810/1/9}, \href
  {https://ui.adsabs.harvard.edu/abs/2015ApJ...810....9C} {810, 9}

\bibitem[\protect\citeauthoryear{{Colombo} et~al.,}{{Colombo}
  et~al.}{2014a}]{2014ApJ...784....3C}
{Colombo} D.,  et~al., 2014a, \mn@doi [\apj] {10.1088/0004-637X/784/1/3}, \href
  {http://adsabs.harvard.edu/abs/2014ApJ...784....3C} {784, 3}

\bibitem[\protect\citeauthoryear{{Colombo} et~al.,}{{Colombo}
  et~al.}{2014b}]{2014ApJ...784....4C}
{Colombo} D.,  et~al., 2014b, \mn@doi [\apj] {10.1088/0004-637X/784/1/4}, \href
  {http://adsabs.harvard.edu/abs/2014ApJ...784....4C} {784, 4}

\bibitem[\protect\citeauthoryear{{Colombo} et~al.,}{{Colombo}
  et~al.}{2019}]{2019MNRAS.483.4291C}
{Colombo} D.,  et~al., 2019, \mn@doi [\mnras] {10.1093/mnras/sty3283}, \href
  {https://ui.adsabs.harvard.edu/abs/2019MNRAS.483.4291C} {483, 4291}

\bibitem[\protect\citeauthoryear{{Cox} \& {G{\'o}mez}}{{Cox} \&
  {G{\'o}mez}}{2002}]{2002ApJS..142..261C}
{Cox} D.~P.,  {G{\'o}mez} G.~C.,  2002, \mn@doi [\apjs] {10.1086/341946}, \href
  {http://adsabs.harvard.edu/abs/2002ApJS..142..261C} {142, 261}

\bibitem[\protect\citeauthoryear{{D'Onghia}}{{D'Onghia}}{2015}]{2015ApJ...808L...8D}
{D'Onghia} E.,  2015, \mn@doi [\apjl] {10.1088/2041-8205/808/1/L8}, \href
  {http://adsabs.harvard.edu/abs/2015ApJ...808L...8D} {808, L8}

\bibitem[\protect\citeauthoryear{{D'Onghia}, {Vogelsberger}  \&
  {Hernquist}}{{D'Onghia} et~al.}{2013}]{2013ApJ...766...34D}
{D'Onghia} E.,  {Vogelsberger} M.,   {Hernquist} L.,  2013, \mn@doi [\apj]
  {10.1088/0004-637X/766/1/34}, \href
  {http://adsabs.harvard.edu/abs/2013ApJ...766...34D} {766, 34}

\bibitem[\protect\citeauthoryear{{Dobbs} \& {Baba}}{{Dobbs} \&
  {Baba}}{2014}]{2014PASA...31...35D}
{Dobbs} C.,  {Baba} J.,  2014, \mn@doi [\pasa] {10.1017/pasa.2014.31}, \href
  {http://adsabs.harvard.edu/abs/2014PASA...31...35D} {31, 35}

\bibitem[\protect\citeauthoryear{{Dobbs} \& {Bonnell}}{{Dobbs} \&
  {Bonnell}}{2006}]{2006MNRAS.367..873D}
{Dobbs} C.~L.,  {Bonnell} I.~A.,  2006, \mn@doi [\mnras]
  {10.1111/j.1365-2966.2006.10146.x}, \href
  {http://adsabs.harvard.edu/abs/2006MNRAS.367..873D} {367, 873}

\bibitem[\protect\citeauthoryear{{Dobbs} \& {Bonnell}}{{Dobbs} \&
  {Bonnell}}{2008}]{2008MNRAS.385.1893D}
{Dobbs} C.~L.,  {Bonnell} I.~A.,  2008, \mn@doi [\mnras]
  {10.1111/j.1365-2966.2008.12995.x}, \href
  {http://adsabs.harvard.edu/abs/2008MNRAS.385.1893D} {385, 1893}

\bibitem[\protect\citeauthoryear{{Dobbs} \& {Pringle}}{{Dobbs} \&
  {Pringle}}{2010}]{2010MNRAS.409..396D}
{Dobbs} C.~L.,  {Pringle} J.~E.,  2010, \mn@doi [\mnras]
  {10.1111/j.1365-2966.2010.17323.x}, \href
  {http://adsabs.harvard.edu/abs/2010MNRAS.409..396D} {409, 396}

\bibitem[\protect\citeauthoryear{{Dobbs}, {Theis}, {Pringle}  \&
  {Bate}}{{Dobbs} et~al.}{2010}]{2010MNRAS.403..625D}
{Dobbs} C.~L.,  {Theis} C.,  {Pringle} J.~E.,   {Bate} M.~R.,  2010, \mn@doi
  [\mnras] {10.1111/j.1365-2966.2009.16161.x}, \href
  {http://adsabs.harvard.edu/abs/2010MNRAS.403..625D} {403, 625}

\bibitem[\protect\citeauthoryear{{Dobbs}, {Burkert}  \& {Pringle}}{{Dobbs}
  et~al.}{2011}]{2011MNRAS.417.1318D}
{Dobbs} C.~L.,  {Burkert} A.,   {Pringle} J.~E.,  2011, \mn@doi [\mnras]
  {10.1111/j.1365-2966.2011.19346.x}, \href
  {http://adsabs.harvard.edu/abs/2011MNRAS.417.1318D} {417, 1318}

\bibitem[\protect\citeauthoryear{{Dobbs}, {Pringle}  \&
  {Duarte-Cabral}}{{Dobbs} et~al.}{2015}]{2015MNRAS.446.3608D}
{Dobbs} C.~L.,  {Pringle} J.~E.,   {Duarte-Cabral} A.,  2015, \mn@doi [\mnras]
  {10.1093/mnras/stu2319}, \href
  {http://adsabs.harvard.edu/abs/2015MNRAS.446.3608D} {446, 3608}

\bibitem[\protect\citeauthoryear{{Dobbs}, {Pettitt}, {Corbelli}  \&
  {Pringle}}{{Dobbs} et~al.}{2018}]{2018MNRAS.478.3793D}
{Dobbs} C.~L.,  {Pettitt} A.~R.,  {Corbelli} E.,   {Pringle} J.~E.,  2018,
  \mn@doi [\mnras] {10.1093/mnras/sty1231}, \href
  {https://ui.adsabs.harvard.edu/abs/2018MNRAS.478.3793D} {478, 3793}

\bibitem[\protect\citeauthoryear{{Dobbs}, {Rosolowsky}, {Pettitt}, {Braine},
  {Corbelli}  \& {Sun}}{{Dobbs} et~al.}{2019}]{2019MNRAS.485.4997D}
{Dobbs} C.~L.,  {Rosolowsky} E.,  {Pettitt} A.~R.,  {Braine} J.,  {Corbelli}
  E.,   {Sun} J.,  2019, \mn@doi [\mnras] {10.1093/mnras/stz674}, \href
  {https://ui.adsabs.harvard.edu/abs/2019MNRAS.485.4997D} {485, 4997}

\bibitem[\protect\citeauthoryear{{Duarte-Cabral} \& {Dobbs}}{{Duarte-Cabral} \&
  {Dobbs}}{2016}]{2016MNRAS.458.3667D}
{Duarte-Cabral} A.,  {Dobbs} C.~L.,  2016, \mn@doi [\mnras]
  {10.1093/mnras/stw469}, \href
  {http://adsabs.harvard.edu/abs/2016MNRAS.458.3667D} {458, 3667}

\bibitem[\protect\citeauthoryear{{Egusa}, {Mentuch Cooper}, {Koda}  \&
  {Baba}}{{Egusa} et~al.}{2017}]{2017MNRAS.465..460E}
{Egusa} F.,  {Mentuch Cooper} E.,  {Koda} J.,   {Baba} J.,  2017, \mn@doi
  [\mnras] {10.1093/mnras/stw2710}, \href
  {http://adsabs.harvard.edu/abs/2017MNRAS.465..460E} {465, 460}

\bibitem[\protect\citeauthoryear{{Eilers}, {Hogg}, {Rix}  \& {Ness}}{{Eilers}
  et~al.}{2019}]{2019ApJ...871..120E}
{Eilers} A.-C.,  {Hogg} D.~W.,  {Rix} H.-W.,   {Ness} M.~K.,  2019, \mn@doi
  [\apj] {10.3847/1538-4357/aaf648}, \href
  {https://ui.adsabs.harvard.edu/abs/2019ApJ...871..120E} {871, 120}

\bibitem[\protect\citeauthoryear{{Foyle}, {Rix}, {Dobbs}, {Leroy}  \&
  {Walter}}{{Foyle} et~al.}{2011}]{2011ApJ...735..101F}
{Foyle} K.,  {Rix} H.-W.,  {Dobbs} C.~L.,  {Leroy} A.~K.,   {Walter} F.,  2011,
  \mn@doi [\apj] {10.1088/0004-637X/735/2/101}, \href
  {https://ui.adsabs.harvard.edu/abs/2011ApJ...735..101F} {735, 101}

\bibitem[\protect\citeauthoryear{{Fujii}, {Baba}, {Saitoh}, {Makino}, {Kokubo}
  \& {Wada}}{{Fujii} et~al.}{2011}]{2011ApJ...730..109F}
{Fujii} M.~S.,  {Baba} J.,  {Saitoh} T.~R.,  {Makino} J.,  {Kokubo} E.,
  {Wada} K.,  2011, \mn@doi [\apj] {10.1088/0004-637X/730/2/109}, \href
  {http://adsabs.harvard.edu/abs/2011ApJ...730..109F} {730, 109}

\bibitem[\protect\citeauthoryear{{Fujimoto}}{{Fujimoto}}{1968}]{1968IAUS...29..453F}
{Fujimoto} M.,  1968, in {Arakelyan} A.,  ed.,  IAU Symposium Vol. 29,
  Non-stable Phenomena in Galaxies. p.~453

\bibitem[\protect\citeauthoryear{{Fujimoto}, {Tasker}, {Wakayama}  \&
  {Habe}}{{Fujimoto} et~al.}{2014}]{2014MNRAS.439..936F}
{Fujimoto} Y.,  {Tasker} E.~J.,  {Wakayama} M.,   {Habe} A.,  2014, \mn@doi
  [\mnras] {10.1093/mnras/stu014}, \href
  {https://ui.adsabs.harvard.edu/abs/2014MNRAS.439..936F} {439, 936}

\bibitem[\protect\citeauthoryear{{Gaia Collaboration} et~al.,}{{Gaia
  Collaboration} et~al.}{2018}]{2018A&A...616A..11G}
{Gaia Collaboration} et~al., 2018, \mn@doi [\aap]
  {10.1051/0004-6361/201832865}, \href
  {https://ui.adsabs.harvard.edu/\#abs/2018A&A...616A..11G} {616, A11}

\bibitem[\protect\citeauthoryear{{Gerhard}}{{Gerhard}}{2011}]{2011MSAIS..18..185G}
{Gerhard} O.,  2011, Memorie della Societa Astronomica Italiana Supplementi,
  \href {http://adsabs.harvard.edu/abs/2011MSAIS..18..185G} {18, 185}

\bibitem[\protect\citeauthoryear{{Grand}, {Kawata}  \& {Cropper}}{{Grand}
  et~al.}{2012}]{2012MNRAS.426..167G}
{Grand} R.~J.~J.,  {Kawata} D.,   {Cropper} M.,  2012, \mn@doi [\mnras]
  {10.1111/j.1365-2966.2012.21733.x}, \href
  {http://adsabs.harvard.edu/abs/2012MNRAS.426..167G} {426, 167}

\bibitem[\protect\citeauthoryear{{Grand}, {Kawata}  \& {Cropper}}{{Grand}
  et~al.}{2013}]{2013A&A...553A..77G}
{Grand} R.~J.~J.,  {Kawata} D.,   {Cropper} M.,  2013, \mn@doi [\aap]
  {10.1051/0004-6361/201321308}, \href
  {http://adsabs.harvard.edu/abs/2013A%26A...553A..77G} {553, A77}

\bibitem[\protect\citeauthoryear{{Grand}, {Bovy}, {Kawata}, {Hunt}, {Famaey},
  {Siebert}, {Monari}  \& {Cropper}}{{Grand}
  et~al.}{2015}]{2015MNRAS.453.1867G}
{Grand} R.~J.~J.,  {Bovy} J.,  {Kawata} D.,  {Hunt} J.~A.~S.,  {Famaey} B.,
  {Siebert} A.,  {Monari} G.,   {Cropper} M.,  2015, \mn@doi [\mnras]
  {10.1093/mnras/stv1785}, \href
  {http://adsabs.harvard.edu/abs/2015MNRAS.453.1867G} {453, 1867}

\bibitem[\protect\citeauthoryear{{Guszejnov}, {Grudi{\'c}}, {Offner},
  {Boylan-Kolchin}, {Faucher-Gig{\`e}re}, {Wetzel}, {Benincasa}  \&
  {Loebman}}{{Guszejnov} et~al.}{2020}]{2020MNRAS.492..488G}
{Guszejnov} D.,  {Grudi{\'c}} M.~Y.,  {Offner} S. S.~R.,  {Boylan-Kolchin} M.,
  {Faucher-Gig{\`e}re} C.-A.,  {Wetzel} A.,  {Benincasa} S.~M.,   {Loebman} S.,
   2020, \mn@doi [\mnras] {10.1093/mnras/stz3527}, \href
  {https://ui.adsabs.harvard.edu/abs/2020MNRAS.492..488G} {492, 488}

\bibitem[\protect\citeauthoryear{{Hirota} et~al.,}{{Hirota}
  et~al.}{2018}]{2018PASJ...70...73H}
{Hirota} A.,  et~al., 2018, \mn@doi [\pasj] {10.1093/pasj/psy071}, \href
  {https://ui.adsabs.harvard.edu/abs/2018PASJ...70...73H} {70, 73}

\bibitem[\protect\citeauthoryear{{Hughes} et~al.,}{{Hughes}
  et~al.}{2013}]{2013ApJ...779...46H}
{Hughes} A.,  et~al., 2013, \mn@doi [\apj] {10.1088/0004-637X/779/1/46}, \href
  {http://adsabs.harvard.edu/abs/2013ApJ...779...46H} {779, 46}

\bibitem[\protect\citeauthoryear{{Hunt}, {Hong}, {Bovy}, {Kawata}  \&
  {Grand}}{{Hunt} et~al.}{2018}]{2018MNRAS.481.3794H}
{Hunt} J. A.~S.,  {Hong} J.,  {Bovy} J.,  {Kawata} D.,   {Grand} R. J.~J.,
  2018, \mn@doi [\mnras] {10.1093/mnras/sty2532}, \href
  {https://ui.adsabs.harvard.edu/abs/2018MNRAS.481.3794H} {481, 3794}

\bibitem[\protect\citeauthoryear{{Junqueira}, {Chiappini}, {L{\'e}pine},
  {Minchev}  \& {Santiago}}{{Junqueira} et~al.}{2015}]{2015MNRAS.449.2336J}
{Junqueira} T.~C.,  {Chiappini} C.,  {L{\'e}pine} J.~R.~D.,  {Minchev} I.,
  {Santiago} B.~X.,  2015, \mn@doi [\mnras] {10.1093/mnras/stv464}, \href
  {http://adsabs.harvard.edu/abs/2015MNRAS.449.2336J} {449, 2336}

\bibitem[\protect\citeauthoryear{{Kawata}, {Hunt}, {Grand}, {Pasetto}  \&
  {Cropper}}{{Kawata} et~al.}{2014}]{2014MNRAS.443.2757K}
{Kawata} D.,  {Hunt} J. A.~S.,  {Grand} R. J.~J.,  {Pasetto} S.,   {Cropper}
  M.,  2014, \mn@doi [\mnras] {10.1093/mnras/stu1292}, \href
  {https://ui.adsabs.harvard.edu/abs/2014MNRAS.443.2757K} {443, 2757}

\bibitem[\protect\citeauthoryear{{Keller}, {Wadsley}, {Benincasa}  \&
  {Couchman}}{{Keller} et~al.}{2014}]{2014MNRAS.442.3013K}
{Keller} B.~W.,  {Wadsley} J.,  {Benincasa} S.~M.,   {Couchman} H.~M.~P.,
  2014, \mn@doi [\mnras] {10.1093/mnras/stu1058}, \href
  {http://adsabs.harvard.edu/abs/2014MNRAS.442.3013K} {442, 3013}

\bibitem[\protect\citeauthoryear{{Khoperskov}, {Eremin}, {Khoperskov},
  {Butenko}  \& {Morozov}}{{Khoperskov} et~al.}{2012}]{2012ARep...56...16K}
{Khoperskov} A.~V.,  {Eremin} M.~A.,  {Khoperskov} S.~A.,  {Butenko} M.~A.,
  {Morozov} A.~G.,  2012, \mn@doi [Astronomy Reports]
  {10.1134/S1063772912010039}, \href
  {http://adsabs.harvard.edu/abs/2012ARep...56...16K} {56, 16}

\bibitem[\protect\citeauthoryear{{Kim} \& {Kim}}{{Kim} \&
  {Kim}}{2014}]{2014MNRAS.440..208K}
{Kim} Y.,  {Kim} W.-T.,  2014, \mn@doi [\mnras] {10.1093/mnras/stu276}, \href
  {http://adsabs.harvard.edu/abs/2014MNRAS.440..208K} {440, 208}

\bibitem[\protect\citeauthoryear{{Kim}, {Kim}  \& {Kim}}{{Kim}
  et~al.}{2014}]{2014ApJ...789...68K}
{Kim} W.-T.,  {Kim} Y.,   {Kim} J.-G.,  2014, \mn@doi [\apj]
  {10.1088/0004-637X/789/1/68}, \href
  {http://adsabs.harvard.edu/abs/2014ApJ...789...68K} {789, 68}

\bibitem[\protect\citeauthoryear{{Koda} \& {Wada}}{{Koda} \&
  {Wada}}{2002}]{2002A&A...396..867K}
{Koda} J.,  {Wada} K.,  2002, \mn@doi [\aap] {10.1051/0004-6361:20021461},
  \href {http://adsabs.harvard.edu/abs/2002A%26A...396..867K} {396, 867}

\bibitem[\protect\citeauthoryear{{Koda} et~al.,}{{Koda}
  et~al.}{2009}]{2009ApJ...700L.132K}
{Koda} J.,  et~al., 2009, \mn@doi [\apjl] {10.1088/0004-637X/700/2/L132}, \href
  {http://adsabs.harvard.edu/abs/2009ApJ...700L.132K} {700, L132}

\bibitem[\protect\citeauthoryear{{Koda} et~al.,}{{Koda}
  et~al.}{2012}]{2012ApJ...761...41K}
{Koda} J.,  et~al., 2012, \mn@doi [\apj] {10.1088/0004-637X/761/1/41}, \href
  {http://adsabs.harvard.edu/abs/2012ApJ...761...41K} {761, 41}

\bibitem[\protect\citeauthoryear{{Kreckel} et~al.,}{{Kreckel}
  et~al.}{2018}]{2018ApJ...863L..21K}
{Kreckel} K.,  et~al., 2018, \mn@doi [\apjl] {10.3847/2041-8213/aad77d}, \href
  {https://ui.adsabs.harvard.edu/abs/2018ApJ...863L..21K} {863, L21}

\bibitem[\protect\citeauthoryear{{Larson}}{{Larson}}{1981}]{1981MNRAS.194..809L}
{Larson} R.~B.,  1981, \mnras, \href
  {http://adsabs.harvard.edu/abs/1981MNRAS.194..809L} {194, 809}

\bibitem[\protect\citeauthoryear{{Leroy} et~al.,}{{Leroy}
  et~al.}{2016}]{2016ApJ...831...16L}
{Leroy} A.~K.,  et~al., 2016, \mn@doi [\apj] {10.3847/0004-637X/831/1/16},
  \href {http://adsabs.harvard.edu/abs/2016ApJ...831...16L} {831, 16}

\bibitem[\protect\citeauthoryear{{Lin} \& {Shu}}{{Lin} \&
  {Shu}}{1964}]{1964ApJ...140..646L}
{Lin} C.~C.,  {Shu} F.~H.,  1964, \mn@doi [\apj] {10.1086/147955}, \href
  {http://adsabs.harvard.edu/abs/1964ApJ...140..646L} {140, 646}

\bibitem[\protect\citeauthoryear{{Lynden-Bell} \& {Kalnajs}}{{Lynden-Bell} \&
  {Kalnajs}}{1972}]{1972MNRAS.157....1L}
{Lynden-Bell} D.,  {Kalnajs} A.~J.,  1972, \mn@doi [\mnras]
  {10.1093/mnras/157.1.1}, \href
  {https://ui.adsabs.harvard.edu/abs/1972MNRAS.157....1L} {157, 1}

\bibitem[\protect\citeauthoryear{{Mart{\'\i}nez-Garc{\'\i}a},
  {Gonz{\'a}lez-L{\'o}pezlira}  \& {Bruzual-A}}{{Mart{\'\i}nez-Garc{\'\i}a}
  et~al.}{2009}]{2009ApJ...694..512M}
{Mart{\'\i}nez-Garc{\'\i}a} E.~E.,  {Gonz{\'a}lez-L{\'o}pezlira} R.~A.,
  {Bruzual-A} G.,  2009, \mn@doi [\apj] {10.1088/0004-637X/694/1/512}, \href
  {https://ui.adsabs.harvard.edu/abs/2009ApJ...694..512M} {694, 512}

\bibitem[\protect\citeauthoryear{{Masters} et~al.,}{{Masters}
  et~al.}{2019}]{2019MNRAS.487.1808M}
{Masters} K.~L.,  et~al., 2019, \mn@doi [\mnras] {10.1093/mnras/stz1153}, \href
  {https://ui.adsabs.harvard.edu/abs/2019MNRAS.487.1808M} {487, 1808}

\bibitem[\protect\citeauthoryear{{Mata-Ch{\'a}vez}, {Vel{\'a}zquez},
  {Pichardo}, {Valenzuela}, {Roca-F{\'a}brega}, {Hern{\'a}ndez-Toledo}  \&
  {Aquino-Ort{\'\i}z}}{{Mata-Ch{\'a}vez} et~al.}{2019}]{2019ApJ...876....6M}
{Mata-Ch{\'a}vez} M.~D.,  {Vel{\'a}zquez} H.,  {Pichardo} B.,  {Valenzuela} O.,
   {Roca-F{\'a}brega} S.,  {Hern{\'a}ndez-Toledo} H.,   {Aquino-Ort{\'\i}z} E.,
   2019, \mn@doi [\apj] {10.3847/1538-4357/ab12d4}, \href
  {https://ui.adsabs.harvard.edu/abs/2019ApJ...876....6M} {876, 6}

\bibitem[\protect\citeauthoryear{{Meidt}, {Rand}, {Merrifield}, {Shetty}  \&
  {Vogel}}{{Meidt} et~al.}{2008}]{2008ApJ...688..224M}
{Meidt} S.~E.,  {Rand} R.~J.,  {Merrifield} M.~R.,  {Shetty} R.,   {Vogel}
  S.~N.,  2008, \mn@doi [\apj] {10.1086/591516}, \href
  {http://adsabs.harvard.edu/abs/2008ApJ...688..224M} {688, 224}

\bibitem[\protect\citeauthoryear{{Meidt} et~al.,}{{Meidt}
  et~al.}{2015}]{2015ApJ...806...72M}
{Meidt} S.~E.,  et~al., 2015, \mn@doi [\apj] {10.1088/0004-637X/806/1/72},
  \href {https://ui.adsabs.harvard.edu/abs/2015ApJ...806...72M} {806, 72}

\bibitem[\protect\citeauthoryear{{Mogotsi} \& {Romeo}}{{Mogotsi} \&
  {Romeo}}{2019}]{2019MNRAS.489.3797M}
{Mogotsi} K.~M.,  {Romeo} A.~B.,  2019, \mn@doi [\mnras]
  {10.1093/mnras/stz2370}, \href
  {https://ui.adsabs.harvard.edu/abs/2019MNRAS.489.3797M} {489, 3797}

\bibitem[\protect\citeauthoryear{{Nguyen}, {Pettitt}, {Tasker}  \&
  {Okamoto}}{{Nguyen} et~al.}{2018}]{2018MNRAS.475...27N}
{Nguyen} N.~K.,  {Pettitt} A.~R.,  {Tasker} E.~J.,   {Okamoto} T.,  2018,
  \mn@doi [\mnras] {10.1093/mnras/stx3143}, \href
  {http://adsabs.harvard.edu/abs/2018MNRAS.475...27N} {475, 27}

\bibitem[\protect\citeauthoryear{{Oh}, {Kim}, {Lee}  \& {Kim}}{{Oh}
  et~al.}{2008}]{2008ApJ...683...94O}
{Oh} S.~H.,  {Kim} W.-T.,  {Lee} H.~M.,   {Kim} J.,  2008, \mn@doi [\apj]
  {10.1086/588184}, \href {http://adsabs.harvard.edu/abs/2008ApJ...683...94O}
  {683, 94}

\bibitem[\protect\citeauthoryear{{Pan}, {Fujimoto}, {Tasker}, {Rosolowsky},
  {Colombo}, {Benincasa}  \& {Wadsley}}{{Pan}
  et~al.}{2015}]{2015MNRAS.453.3082P}
{Pan} H.-A.,  {Fujimoto} Y.,  {Tasker} E.~J.,  {Rosolowsky} E.,  {Colombo} D.,
  {Benincasa} S.~M.,   {Wadsley} J.,  2015, \mn@doi [\mnras]
  {10.1093/mnras/stv1843}, \href
  {http://adsabs.harvard.edu/abs/2015MNRAS.453.3082P} {453, 3082}

\bibitem[\protect\citeauthoryear{{Peterken}, {Merrifield},
  {Arag{\'o}n-Salamanca}, {Drory}, {Krawczyk}, {Masters}, {Weijmans}  \&
  {Westfall}}{{Peterken} et~al.}{2019}]{2019NatAs...3..178P}
{Peterken} T.~G.,  {Merrifield} M.~R.,  {Arag{\'o}n-Salamanca} A.,  {Drory} N.,
   {Krawczyk} C.~M.,  {Masters} K.~L.,  {Weijmans} A.-M.,   {Westfall} K.~B.,
  2019, \mn@doi [Nature Astronomy] {10.1038/s41550-018-0627-5}, \href
  {https://ui.adsabs.harvard.edu/abs/2019NatAs...3..178P} {3, 178}

\bibitem[\protect\citeauthoryear{{Pettitt} \& {Wadsley}}{{Pettitt} \&
  {Wadsley}}{2018}]{2018MNRAS.474.5645P}
{Pettitt} A.~R.,  {Wadsley} J.~W.,  2018, \mn@doi [\mnras]
  {10.1093/mnras/stx3129}, \href
  {http://adsabs.harvard.edu/abs/2018MNRAS.474.5645P} {474, 5645}

\bibitem[\protect\citeauthoryear{{Pettitt}, {Dobbs}, {Acreman}  \&
  {Price}}{{Pettitt} et~al.}{2014}]{2014arXiv1406.4150P}
{Pettitt} A.~R.,  {Dobbs} C.~L.,  {Acreman} D.~M.,   {Price} D.~J.,  2014,
  \mnras, \href {http://adsabs.harvard.edu/abs/2014arXiv1406.4150P} {444, 919}

\bibitem[\protect\citeauthoryear{{Pettitt}, {Dobbs}, {Acreman}  \&
  {Bate}}{{Pettitt} et~al.}{2015}]{2015MNRAS.449.3911P}
{Pettitt} A.~R.,  {Dobbs} C.~L.,  {Acreman} D.~M.,   {Bate} M.~R.,  2015,
  \mn@doi [\mnras] {10.1093/mnras/stv600}, \href
  {http://adsabs.harvard.edu/abs/2015MNRAS.449.3911P} {449, 3911}

\bibitem[\protect\citeauthoryear{{Pettitt}, {Tasker}  \& {Wadsley}}{{Pettitt}
  et~al.}{2016}]{2016MNRAS.458.3990P}
{Pettitt} A.~R.,  {Tasker} E.~J.,   {Wadsley} J.~W.,  2016, \mn@doi [\mnras]
  {10.1093/mnras/stw588}, \href
  {http://adsabs.harvard.edu/abs/2016MNRAS.458.3990P} {458, 3990}

\bibitem[\protect\citeauthoryear{{Pettitt}, {Tasker}, {Wadsley}, {Keller}  \&
  {Benincasa}}{{Pettitt} et~al.}{2017}]{2017MNRAS.468.4189P}
{Pettitt} A.~R.,  {Tasker} E.~J.,  {Wadsley} J.~W.,  {Keller} B.~W.,
  {Benincasa} S.~M.,  2017, \mn@doi [\mnras] {10.1093/mnras/stx736}, \href
  {http://adsabs.harvard.edu/abs/2017MNRAS.468.4189P} {468, 4189}

\bibitem[\protect\citeauthoryear{{Pettitt}, {Egusa}, {Dobbs}, {Tasker},
  {Fujimoto}  \& {Habe}}{{Pettitt} et~al.}{2018}]{2018MNRAS.480.3356P}
{Pettitt} A.~R.,  {Egusa} F.,  {Dobbs} C.~L.,  {Tasker} E.~J.,  {Fujimoto} Y.,
   {Habe} A.,  2018, \mn@doi [\mnras] {10.1093/mnras/sty2040}, \href
  {https://ui.adsabs.harvard.edu/abs/2018MNRAS.480.3356P} {480, 3356}

\bibitem[\protect\citeauthoryear{{Pettitt}, {Ragan}  \& {Smith}}{{Pettitt}
  et~al.}{2020}]{2020MNRAS.491.2162P}
{Pettitt} A.~R.,  {Ragan} S.~E.,   {Smith} M.~C.,  2020, \mn@doi [\mnras]
  {10.1093/mnras/stz3155}, \href
  {https://ui.adsabs.harvard.edu/abs/2020MNRAS.491.2162P} {491, 2162}

\bibitem[\protect\citeauthoryear{{Pontzen}, {Ro{\v s}kar}, {Stinson}  \&
  {Woods}}{{Pontzen} et~al.}{2013}]{2013ascl.soft05002P}
{Pontzen} A.,  {Ro{\v s}kar} R.,  {Stinson} G.,   {Woods} R.,  2013, {pynbody:
  N-Body/SPH analysis for python}, Astrophysics Source Code Library (\mn@eprint
  {ascl} {1305.002})

\bibitem[\protect\citeauthoryear{{Pour-Imani}, {Kennefick}, {Kennefick},
  {Davis}, {Shields}  \& {Shameer Abdeen}}{{Pour-Imani}
  et~al.}{2016}]{2016ApJ...827L...2P}
{Pour-Imani} H.,  {Kennefick} D.,  {Kennefick} J.,  {Davis} B.~L.,  {Shields}
  D.~W.,   {Shameer Abdeen} M.,  2016, \mn@doi [\apjl]
  {10.3847/2041-8205/827/1/L2}, \href
  {https://ui.adsabs.harvard.edu/abs/2016ApJ...827L...2P} {827, L2}

\bibitem[\protect\citeauthoryear{{Pringle} \& {Dobbs}}{{Pringle} \&
  {Dobbs}}{2019}]{2019MNRAS.490.1470P}
{Pringle} J.~E.,  {Dobbs} C.~L.,  2019, \mn@doi [\mnras]
  {10.1093/mnras/stz2694}, \href
  {https://ui.adsabs.harvard.edu/abs/2019MNRAS.490.1470P} {490, 1470}

\bibitem[\protect\citeauthoryear{{Reid} et~al.,}{{Reid}
  et~al.}{2019}]{2019ApJ...885..131R}
{Reid} M.~J.,  et~al., 2019, \mn@doi [\apj] {10.3847/1538-4357/ab4a11}, \href
  {https://ui.adsabs.harvard.edu/abs/2019ApJ...885..131R} {885, 131}

\bibitem[\protect\citeauthoryear{{Rice}, {Goodman}, {Bergin}, {Beaumont}  \&
  {Dame}}{{Rice} et~al.}{2016}]{2016ApJ...822...52R}
{Rice} T.~S.,  {Goodman} A.~A.,  {Bergin} E.~A.,  {Beaumont} C.,   {Dame}
  T.~M.,  2016, \mn@doi [\apj] {10.3847/0004-637X/822/1/52}, \href
  {https://ui.adsabs.harvard.edu/abs/2016ApJ...822...52R} {822, 52}

\bibitem[\protect\citeauthoryear{{Rigby} et~al.,}{{Rigby}
  et~al.}{2019}]{2019A&A...632A..58R}
{Rigby} A.~J.,  et~al., 2019, \mn@doi [\aap] {10.1051/0004-6361/201935236},
  \href {https://ui.adsabs.harvard.edu/abs/2019A&A...632A..58R} {632, A58}

\bibitem[\protect\citeauthoryear{{Roberts}}{{Roberts}}{1969}]{1969ApJ...158..123R}
{Roberts} W.~W.,  1969, \mn@doi [\apj] {10.1086/150177}, \href
  {http://adsabs.harvard.edu/abs/1969ApJ...158..123R} {158, 123}

\bibitem[\protect\citeauthoryear{{Rosolowsky}}{{Rosolowsky}}{2005}]{2005PASP..117.1403R}
{Rosolowsky} E.,  2005, \mn@doi [\pasp] {10.1086/497582}, \href
  {http://adsabs.harvard.edu/abs/2005PASP..117.1403R} {117, 1403}

\bibitem[\protect\citeauthoryear{{Sanders} \& {Huntley}}{{Sanders} \&
  {Huntley}}{1976}]{1976ApJ...209...53S}
{Sanders} R.~H.,  {Huntley} J.~M.,  1976, \mn@doi [\apj] {10.1086/154692},
  \href {https://ui.adsabs.harvard.edu/abs/1976ApJ...209...53S} {209, 53}

\bibitem[\protect\citeauthoryear{{Schinnerer} et~al.,}{{Schinnerer}
  et~al.}{2017}]{2017ApJ...836...62S}
{Schinnerer} E.,  et~al., 2017, \mn@doi [\apj] {10.3847/1538-4357/836/1/62},
  \href {http://adsabs.harvard.edu/abs/2017ApJ...836...62S} {836, 62}

\bibitem[\protect\citeauthoryear{{Schinnerer} et~al.,}{{Schinnerer}
  et~al.}{2019}]{2019ApJ...887...49S}
{Schinnerer} E.,  et~al., 2019, \mn@doi [\apj] {10.3847/1538-4357/ab50c2},
  \href {https://ui.adsabs.harvard.edu/abs/2019ApJ...887...49S} {887, 49}

\bibitem[\protect\citeauthoryear{{Sellwood}}{{Sellwood}}{2011}]{2011MNRAS.410.1637S}
{Sellwood} J.~A.,  2011, \mn@doi [\mnras] {10.1111/j.1365-2966.2010.17545.x},
  \href {http://adsabs.harvard.edu/abs/2011MNRAS.410.1637S} {410, 1637}

\bibitem[\protect\citeauthoryear{{Sellwood}}{{Sellwood}}{2012}]{2012ApJ...751...44S}
{Sellwood} J.~A.,  2012, \mn@doi [\apj] {10.1088/0004-637X/751/1/44}, \href
  {https://ui.adsabs.harvard.edu/abs/2012ApJ...751...44S} {751, 44}

\bibitem[\protect\citeauthoryear{{Sellwood} \& {Binney}}{{Sellwood} \&
  {Binney}}{2002}]{2002MNRAS.336..785S}
{Sellwood} J.~A.,  {Binney} J.~J.,  2002, \mn@doi [\mnras]
  {10.1046/j.1365-8711.2002.05806.x}, \href
  {http://adsabs.harvard.edu/abs/2002MNRAS.336..785S} {336, 785}

\bibitem[\protect\citeauthoryear{{Sellwood} \& {Carlberg}}{{Sellwood} \&
  {Carlberg}}{1984}]{1984ApJ...282...61S}
{Sellwood} J.~A.,  {Carlberg} R.~G.,  1984, \mn@doi [\apj] {10.1086/162176},
  \href {http://adsabs.harvard.edu/abs/1984ApJ...282...61S} {282, 61}

\bibitem[\protect\citeauthoryear{{Sellwood} \& {Carlberg}}{{Sellwood} \&
  {Carlberg}}{2014}]{2014ApJ...785..137S}
{Sellwood} J.~A.,  {Carlberg} R.~G.,  2014, \mn@doi [\apj]
  {10.1088/0004-637X/785/2/137}, \href
  {http://adsabs.harvard.edu/abs/2014ApJ...785..137S} {785, 137}

\bibitem[\protect\citeauthoryear{{Sellwood} \& {Carlberg}}{{Sellwood} \&
  {Carlberg}}{2019}]{2019MNRAS.489..116S}
{Sellwood} J.~A.,  {Carlberg} R.~G.,  2019, \mn@doi [\mnras]
  {10.1093/mnras/stz2132}, \href
  {https://ui.adsabs.harvard.edu/abs/2019MNRAS.489..116S} {489, 116}

\bibitem[\protect\citeauthoryear{{Sellwood}, {Trick}, {Carlberg}, {Coronado}
  \& {Rix}}{{Sellwood} et~al.}{2019}]{2019MNRAS.484.3154S}
{Sellwood} J.~A.,  {Trick} W.~H.,  {Carlberg} R.~G.,  {Coronado} J.,   {Rix}
  H.-W.,  2019, \mn@doi [\mnras] {10.1093/mnras/stz140}, \href
  {http://adsabs.harvard.edu/abs/2019MNRAS.484.3154S} {484, 3154}

\bibitem[\protect\citeauthoryear{{Shabani} et~al.,}{{Shabani}
  et~al.}{2018}]{2018MNRAS.478.3590S}
{Shabani} F.,  et~al., 2018, \mn@doi [\mnras] {10.1093/mnras/sty1277}, \href
  {https://ui.adsabs.harvard.edu/abs/2018MNRAS.478.3590S} {478, 3590}

\bibitem[\protect\citeauthoryear{{Shen}, {Wadsley}  \& {Stinson}}{{Shen}
  et~al.}{2010}]{2010MNRAS.407.1581S}
{Shen} S.,  {Wadsley} J.,   {Stinson} G.,  2010, \mn@doi [\mnras]
  {10.1111/j.1365-2966.2010.17047.x}, \href
  {http://adsabs.harvard.edu/abs/2010MNRAS.407.1581S} {407, 1581}

\bibitem[\protect\citeauthoryear{{Shetty} \& {Ostriker}}{{Shetty} \&
  {Ostriker}}{2006}]{2006ApJ...647..997S}
{Shetty} R.,  {Ostriker} E.~C.,  2006, \mn@doi [\apj] {10.1086/505594}, \href
  {http://adsabs.harvard.edu/abs/2006ApJ...647..997S} {647, 997}

\bibitem[\protect\citeauthoryear{{Shetty}, {Vogel}, {Ostriker}  \&
  {Teuben}}{{Shetty} et~al.}{2007}]{2007ApJ...665.1138S}
{Shetty} R.,  {Vogel} S.~N.,  {Ostriker} E.~C.,   {Teuben} P.~J.,  2007,
  \mn@doi [\apj] {10.1086/520037}, \href
  {https://ui.adsabs.harvard.edu/abs/2007ApJ...665.1138S} {665, 1138}

\bibitem[\protect\citeauthoryear{{Shu}}{{Shu}}{2016}]{2016ARA&A..54..667S}
{Shu} F.~H.,  2016, \mn@doi [\araa] {10.1146/annurev-astro-081915-023426},
  \href {https://ui.adsabs.harvard.edu/abs/2016ARA&A..54..667S} {54, 667}

\bibitem[\protect\citeauthoryear{{Shu}, {Milione}  \& {Roberts}}{{Shu}
  et~al.}{1973}]{1973ApJ...183..819S}
{Shu} F.~H.,  {Milione} V.,   {Roberts} Jr. W.~W.,  1973, \mn@doi [\apj]
  {10.1086/152270}, \href
  {https://ui.adsabs.harvard.edu/abs/1973ApJ...183..819S} {183, 819}

\bibitem[\protect\citeauthoryear{{Siebert} et~al.,}{{Siebert}
  et~al.}{2012}]{2012MNRAS.425.2335S}
{Siebert} A.,  et~al., 2012, \mn@doi [\mnras]
  {10.1111/j.1365-2966.2012.21638.x}, \href
  {https://ui.adsabs.harvard.edu/abs/2012MNRAS.425.2335S} {425, 2335}

\bibitem[\protect\citeauthoryear{{Smith} et~al.,}{{Smith}
  et~al.}{2020}]{2020MNRAS.492.1594S}
{Smith} R.~J.,  et~al., 2020, \mn@doi [\mnras] {10.1093/mnras/stz3328}, \href
  {https://ui.adsabs.harvard.edu/abs/2020MNRAS.492.1594S} {492, 1594}

\bibitem[\protect\citeauthoryear{{Stinson}, {Seth}, {Katz}, {Wadsley},
  {Governato}  \& {Quinn}}{{Stinson} et~al.}{2006}]{2006MNRAS.373.1074S}
{Stinson} G.,  {Seth} A.,  {Katz} N.,  {Wadsley} J.,  {Governato} F.,   {Quinn}
  T.,  2006, \mn@doi [\mnras] {10.1111/j.1365-2966.2006.11097.x}, \href
  {http://adsabs.harvard.edu/abs/2006MNRAS.373.1074S} {373, 1074}

\bibitem[\protect\citeauthoryear{{Stinson}, {Brook}, {Macci{\`o}}, {Wadsley},
  {Quinn}  \& {Couchman}}{{Stinson} et~al.}{2013}]{2013MNRAS.428..129S}
{Stinson} G.~S.,  {Brook} C.,  {Macci{\`o}} A.~V.,  {Wadsley} J.,  {Quinn}
  T.~R.,   {Couchman} H.~M.~P.,  2013, \mn@doi [\mnras] {10.1093/mnras/sts028},
  \href {http://adsabs.harvard.edu/abs/2013MNRAS.428..129S} {428, 129}

\bibitem[\protect\citeauthoryear{{Struck}, {Dobbs}  \& {Hwang}}{{Struck}
  et~al.}{2011}]{2011MNRAS.414.2498S}
{Struck} C.,  {Dobbs} C.~L.,   {Hwang} J.-S.,  2011, \mn@doi [\mnras]
  {10.1111/j.1365-2966.2011.18568.x}, \href
  {http://adsabs.harvard.edu/abs/2011MNRAS.414.2498S} {414, 2498}

\bibitem[\protect\citeauthoryear{{Sun} et~al.,}{{Sun}
  et~al.}{2018}]{2018ApJ...860..172S}
{Sun} J.,  et~al., 2018, \mn@doi [\apj] {10.3847/1538-4357/aac326}, \href
  {https://ui.adsabs.harvard.edu/abs/2018ApJ...860..172S} {860, 172}

\bibitem[\protect\citeauthoryear{{Tasker}, {Wadsley}  \& {Pudritz}}{{Tasker}
  et~al.}{2015}]{2015ApJ...801...33T}
{Tasker} E.~J.,  {Wadsley} J.,   {Pudritz} R.,  2015, \mn@doi [\apj]
  {10.1088/0004-637X/801/1/33}, \href
  {http://adsabs.harvard.edu/abs/2015ApJ...801...33T} {801, 33}

\bibitem[\protect\citeauthoryear{{Teuben}}{{Teuben}}{1995}]{1995ASPC...77..398T}
{Teuben} P.,  1995, in {Shaw} R.~A.,  {Payne} H.~E.,   {Hayes} J.~J.~E.,  eds,
  Astronomical Society of the Pacific Conference Series Vol. 77, Astronomical
  Data Analysis Software and Systems IV. p.~398

\bibitem[\protect\citeauthoryear{{Toomre} \& {Toomre}}{{Toomre} \&
  {Toomre}}{1972}]{1972ApJ...178..623T}
{Toomre} A.,  {Toomre} J.,  1972, \mn@doi [\apj] {10.1086/151823}, \href
  {http://adsabs.harvard.edu/abs/1972ApJ...178..623T} {178, 623}

\bibitem[\protect\citeauthoryear{{Venturi} et~al.,}{{Venturi}
  et~al.}{2018}]{2018A&A...619A..74V}
{Venturi} G.,  et~al., 2018, \mn@doi [\aap] {10.1051/0004-6361/201833668},
  \href {https://ui.adsabs.harvard.edu/abs/2018A&A...619A..74V} {619, A74}

\bibitem[\protect\citeauthoryear{{Wada} \& {Koda}}{{Wada} \&
  {Koda}}{2001}]{2001PASJ...53.1163W}
{Wada} K.,  {Koda} J.,  2001, \pasj, \href
  {http://adsabs.harvard.edu/abs/2001PASJ...53.1163W} {53, 1163}

\bibitem[\protect\citeauthoryear{{Wada} \& {Koda}}{{Wada} \&
  {Koda}}{2004}]{2004MNRAS.349..270W}
{Wada} K.,  {Koda} J.,  2004, \mn@doi [\mnras]
  {10.1111/j.1365-2966.2004.07484.x}, \href
  {http://adsabs.harvard.edu/abs/2004MNRAS.349..270W} {349, 270}

\bibitem[\protect\citeauthoryear{{Wada}, {Baba}  \& {Saitoh}}{{Wada}
  et~al.}{2011}]{2011ApJ...735....1W}
{Wada} K.,  {Baba} J.,   {Saitoh} T.~R.,  2011, \mn@doi [\apj]
  {10.1088/0004-637X/735/1/1}, \href
  {http://adsabs.harvard.edu/abs/2011ApJ...735....1W} {735, 1}

\bibitem[\protect\citeauthoryear{{Wadsley}, {Keller}  \& {Quinn}}{{Wadsley}
  et~al.}{2017}]{2017MNRAS.471.2357W}
{Wadsley} J.~W.,  {Keller} B.~W.,   {Quinn} T.~R.,  2017, \mn@doi [\mnras]
  {10.1093/mnras/stx1643}, \href
  {http://adsabs.harvard.edu/abs/2017MNRAS.471.2357W} {471, 2357}

\bibitem[\protect\citeauthoryear{{Yu} \& {Ho}}{{Yu} \&
  {Ho}}{2018}]{2018ApJ...869...29Y}
{Yu} S.-Y.,  {Ho} L.~C.,  2018, \mn@doi [\apj] {10.3847/1538-4357/aaeacd},
  \href {https://ui.adsabs.harvard.edu/abs/2018ApJ...869...29Y} {869, 29}

\bibitem[\protect\citeauthoryear{{Yun}}{{Yun}}{1999}]{1999IAUS..186...81Y}
{Yun} M.~S.,  1999, in {Barnes} J.~E.,  {Sanders} D.~B.,  eds,  IAU Symposium
  Vol. 186, Galaxy Interactions at Low and High Redshift. Kluwer Academic
  Publishers, Dordrech., p.~81

\makeatother
\end{thebibliography}

\appendix

\section[]{Supplementary data}
\label{AppA}

\begin{figure}
\begin{centering}
\includegraphics[trim =0mm 0mm 0mm 0mm,width=90mm]{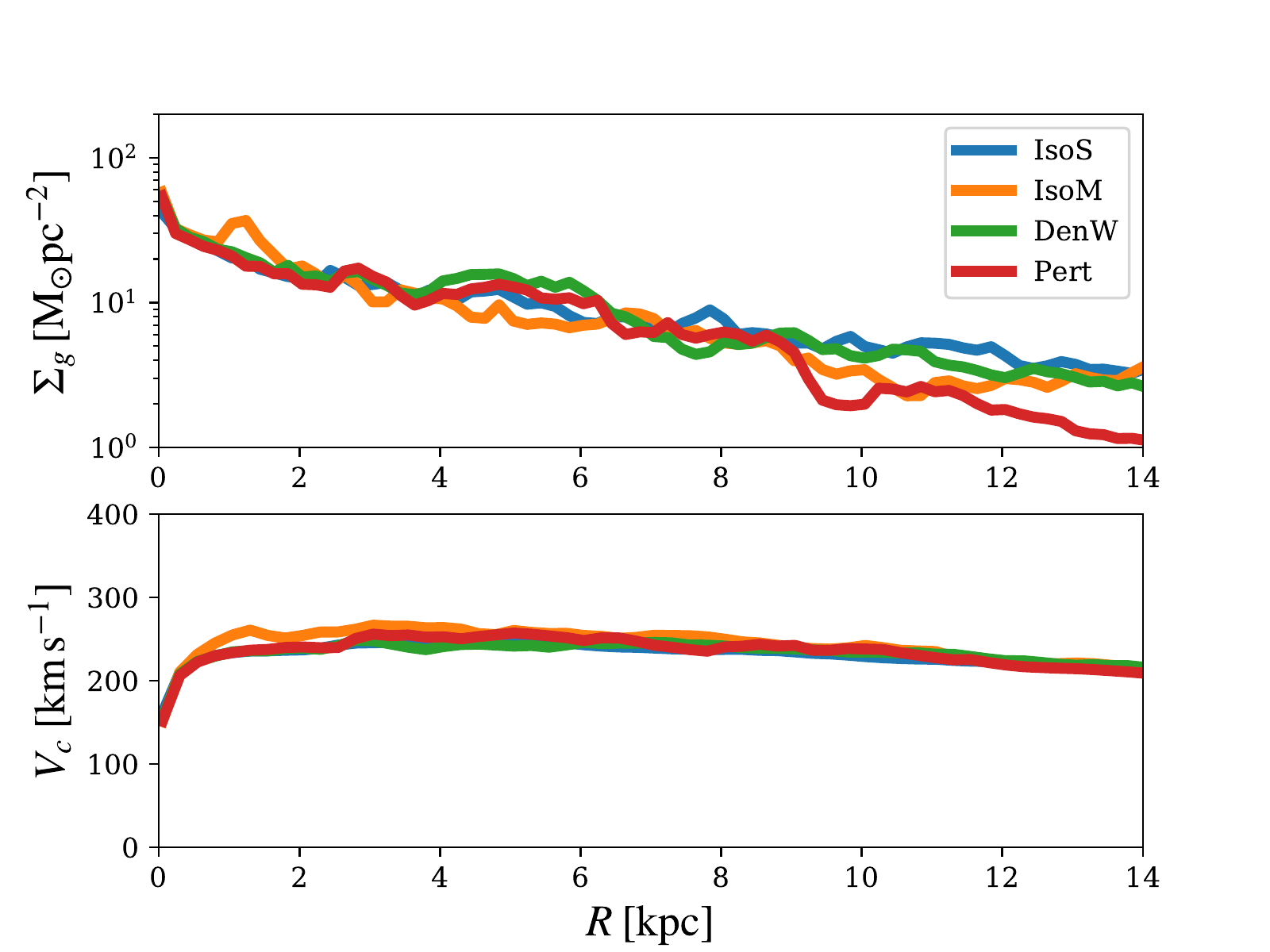}
\caption{Azimuthally averaged gas surface densities and rotation curves for each model as a function of galactic radius.}
\label{VcSd}
\end{centering}
\end{figure}

Figure\;\ref{VcSd} shows the rotation curves and gas surface densities for each of the models. All models have effectively the same profiles with only minor differences. $\Sigma_g$ is slightly smaller for Pert at outer radii due to tidal stripping by the companion. IsoM exhibits a slightly higher $V_c$ in the inner disc, which is due to the re-allocation of mass to the disc profile, which is more centrally concentrated than the halo from which mass has been subtracted.

\begin{figure}
\begin{centering}
\includegraphics[trim =0mm 10mm 0mm 0mm,width=80mm]{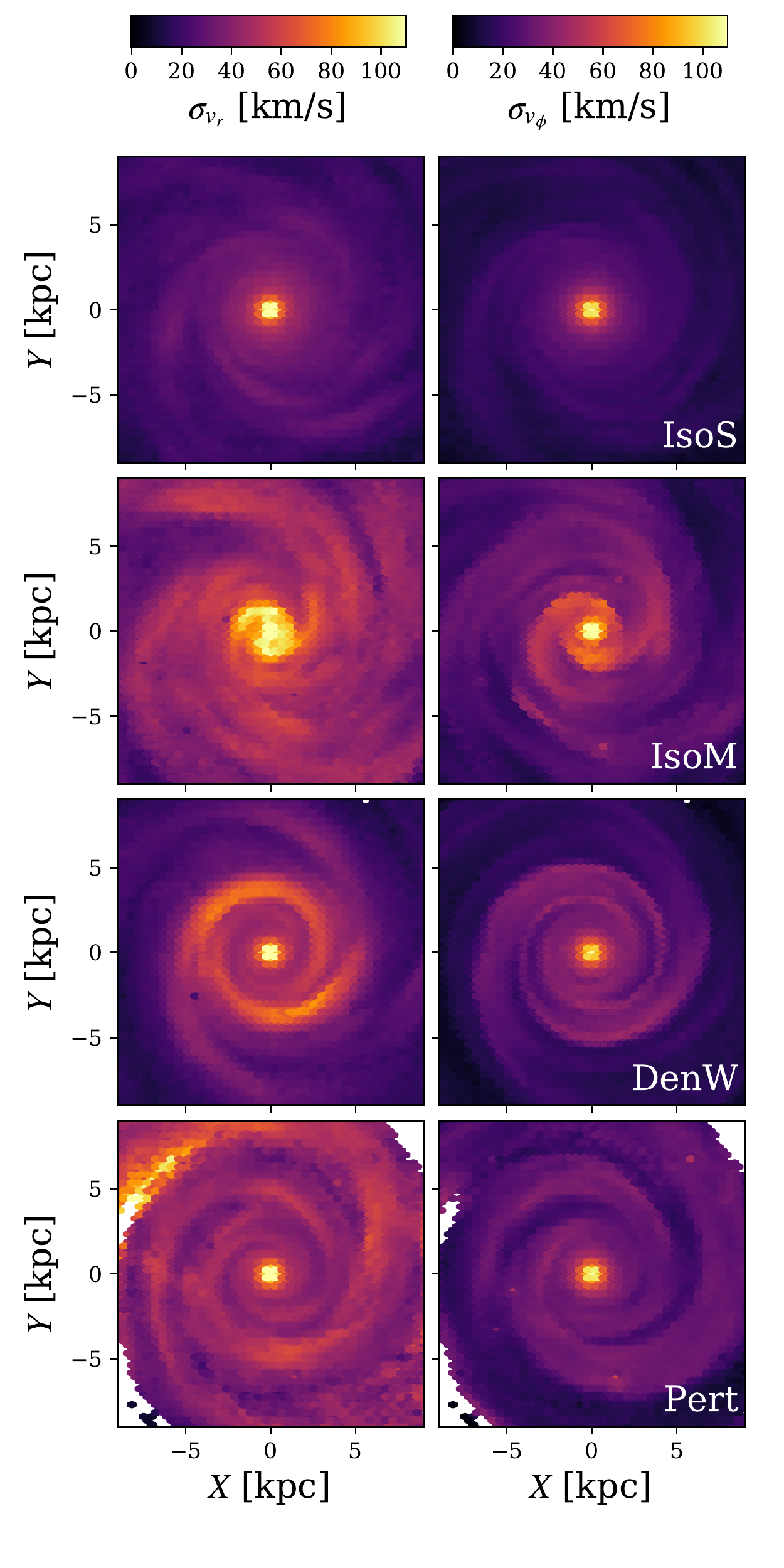}
\caption{Radial (left) and azimuthal (right) stellar velocity dispersion maps for each model. From top: IsoS, IsoM, DenW, Pert.}
\label{starsigma}
\end{centering}
\end{figure}

Fig.\;\ref{starsigma} shows the stellar velocity dispersions (radially and azimuthally) in the four different models. Each disc clearly has a different response. The radial dispersion (left column) in particular shows interesting differences in the grand-design discs compared to the flocculent disc (top row). In the heavy disc model (second row) the disc is heated up across all radii. For the disc experiencing a rigid spiral wave (third row) the heating is limited to a small radial range very close to the ILR. In the tidally perturbed disc (bottom row) the outer disc region is most efficiently heated, and is so up to similar levels as the inner regions of IsoM.

\begin{figure}
\begin{centering}
\includegraphics[trim =0mm 0mm 0mm 0mm,width=90mm]{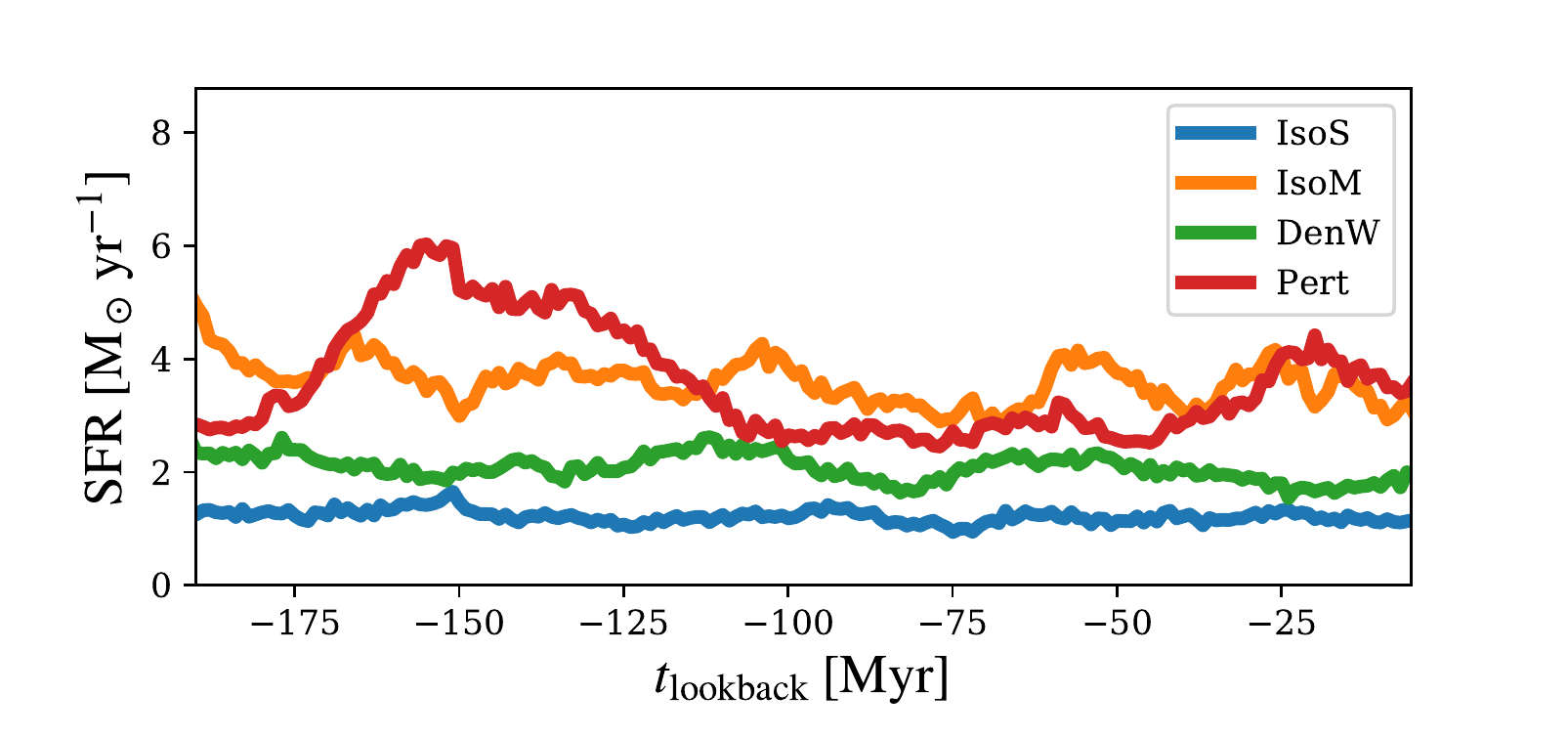}
\caption{Star formation history as a function of time leading up to the point of GMC/ISM analysis presented in this study.}
\label{SFH}
\end{centering}
\end{figure}

Fig.\;\ref{SFH} shows the star formation history across roughly half a dynamical time from the time-frame where the analysis in this study is performed, a time-frame that includes the isolated period of the Pert disc before perigalacticon passage of the perturbed at around $-150$\,Myr. Each isolated model is separated by approximately $1\rm \;M_\odot\,yr^{-1}$, with IsoS being the lowest and IsoM being the highest. The interacting model reaches the highest rates of star formation (coinciding with just after peri-galacticon passage, see P17) but only for a small window until dropping down to levels similar to the other grand design spiral models afterwards.

\section{Histograms of cloud properties}
\label{AppB}
\begin{figure}
\begin{centering}
\includegraphics[trim =0mm 0mm 0mm 0mm,width=80mm]{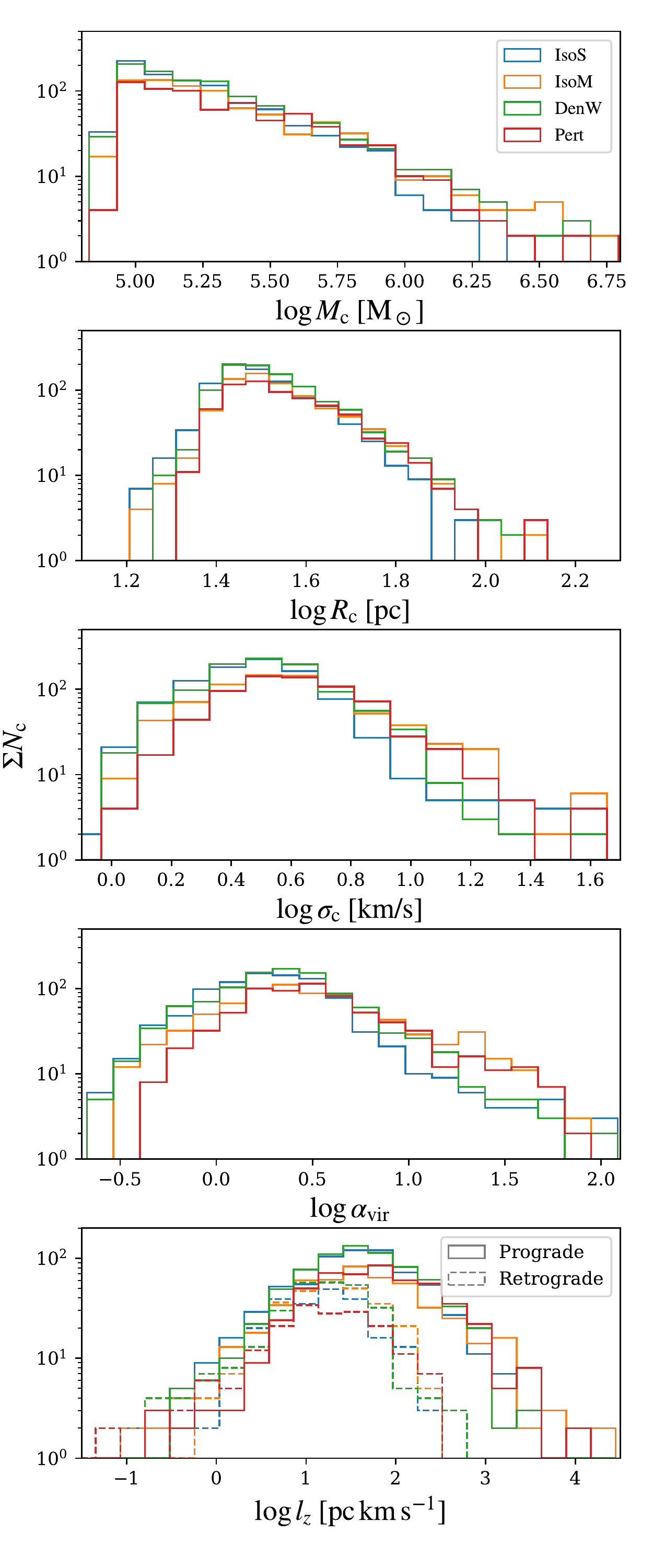}
\caption{Various properties of GMCs in each different model of spiral arm generation using the entire disc population. Top to bottom we show the clouds mass, radius, velocity dispersion, virial parameter and angular momentum in the $z$-direction, the latter shown for both pro and retrograde rotations in solid and dashed lines. See the discussion of Figure\;\ref{ModelHistCompV} in the main text.}
\label{ModelHistComp}
\end{centering}
\end{figure}

\begin{figure}
\begin{centering}
\includegraphics[trim =0mm 0mm 0mm 0mm,width=80mm]{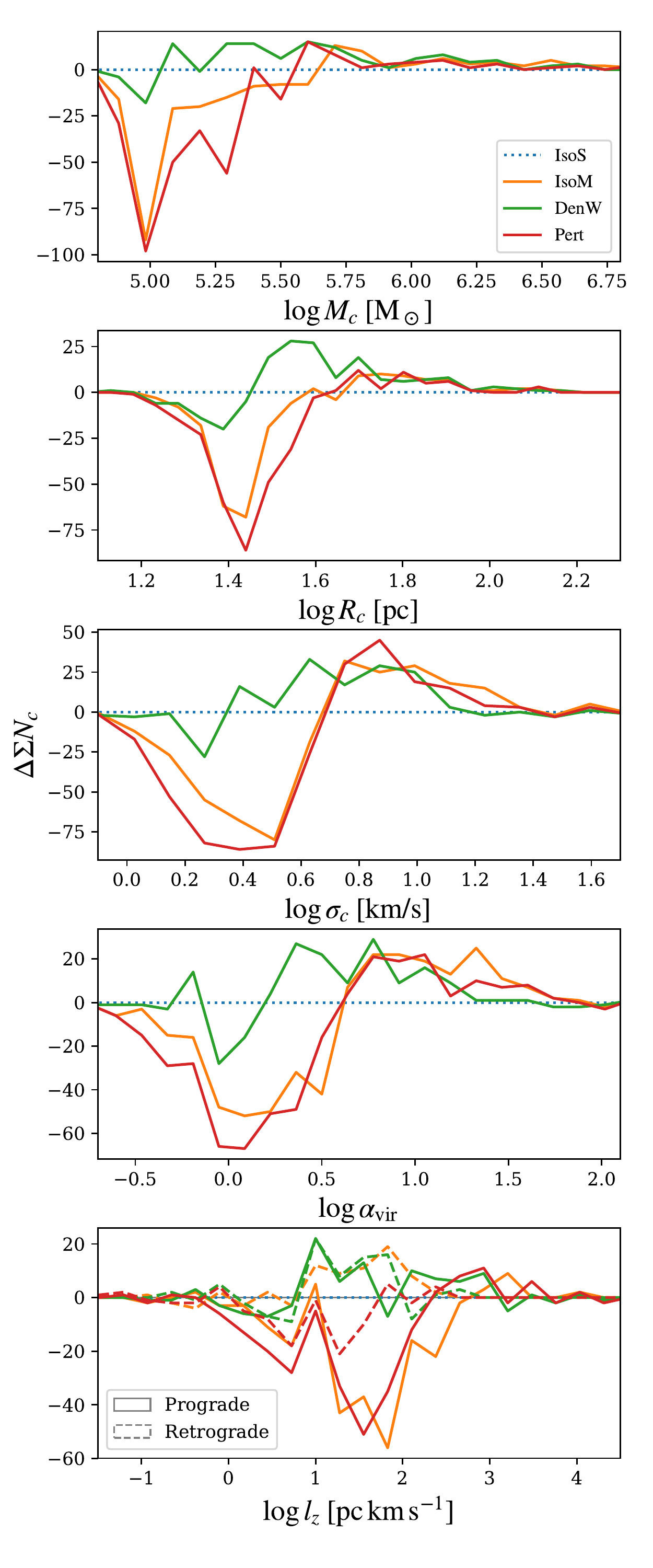}
\caption{The same data as shown in Fig.\;\ref{ModelHistComp} but with each model subtracted from the IsoS model to highlight changes in the distributions compared to the flocculent disc. Note the differences in scales in the $y$-axis in each panel.}
\label{ModelHistCompD}
\end{centering}
\end{figure}

In Figure\;\ref{ModelHistComp} we present a histogram representation of the same data shown in the violin plots in Figure\;\ref{ModelHistCompV}. These histograms show the true distribution of the data, without the inherent kernel weighting of the violin plots. The trends seen are exactly the same as the plots in the main text. In Figure\;\ref{ModelHistCompD} we plot the differences in the distributions for each model relative to IsoS, which highlights the impact of the different grand-design perturbations on the original galactic cloud population.

\bsp
\label{lastpage}
\end{document}